\newlength{\intwidth}

\documentclass{jfm}
\usepackage{amssymb}

\usepackage{array}
\usepackage{amsmath}
\usepackage{latexsym}
\usepackage{bezier}
\usepackage{psfrag,graphicx}
\usepackage{bm}
\usepackage{float}
\usepackage{multirow}
\usepackage{mathrsfs}
\usepackage{color}

\begin{document}


\title[\textcolor{black}{High-speed shear driven dynamos. Part 2. Numerical analysis}]{\textcolor{black}{High-speed shear driven dynamos. Part 2. Numerical analysis}}
\author{Kengo Deguchi}
\affiliation{
School of Mathematical Sciences, Monash University, VIC 3800, Australia
}

\maketitle

\begin{abstract}
This paper aims to numerically verify the large Reynolds number asymptotic theory of magneto-hydrodynamic \textcolor{black}{(MHD)} flows proposed in the companion paper Deguchi \textcolor{black}{(2019)}. 
\textcolor{black}{To avoid any} complexity associated with \textcolor{black}{the} chaotic nature of turbulence and flow geometry, nonlinear steady solutions of the viscous-resistive magneto-hydrodynamic equations in plane Couette flow \textcolor{black}{have been utilized.} 
Two classes of \textcolor{black}{nonlinear MHD} states, which convert kinematic energy to magnetic energy effectively, \textcolor{black}{have been determined}.
The first class of \textcolor{black}{nonlinear states} can be obtained when a small spanwise uniform magnetic field is applied to the known hydrodynamic solution branch of \textcolor{black}{the} plane Couette flow. 
The \textcolor{black}{nonlinear states} are characterised by the hydrodynamic/magnetic roll-streak and the resonant layer at which strong vorticity and current sheets are observed. 
\textcolor{black}{These} flow features, and the induced strong streamwise magnetic field, are fully consistent with the vortex/Alfv\'en wave interaction theory proposed in Deguchi \textcolor{black}{(2019)}.
When the spanwise uniform magnetic field is switched off, the solutions become purely hydrodynamic.
However, 
\textcolor{black}{the second class of `self-sustained shear driven dynamos' at the zero-external magnetic field limit} can be found by homotopy via the forced states subject to a spanwise uniform current field.
The discovery of \textcolor{black}{the dynamo states} \textcolor{black}{has} motivated the corresponding \textcolor{black}{large Reynolds number} matched asymptotic analysis in Deguchi \textcolor{black}{(2019)}.
Here, the reduced equations \textcolor{black}{derived by} the asymptotic theory \textcolor{black}{have been} solved numerically.
The asymptotic solution \textcolor{black}{provides} remarkably good predictions for the finite Reynolds number dynamo solutions.
\end{abstract}

\section{Introduction}

Our concern is the large Reynolds number development of unstable invariant solutions underpinning \textcolor{black}{the} turbulent magneto-hydrodynamic (MHD) flows. 
In recent years, the dynamical systems theory \textcolor{black}{view} of turbulence has been studied \textcolor{black}{extensively by} fluid dynamics researchers. 
\textcolor{black}{It was discovered by several} researchers that  
unstable invariant solutions of the equations can be regarded as ÔskeletonsÕ on which the chaotic dynamics of turbulence hang. 
Indeed, some solutions reproduce remarkable statistical properties of turbulent flows,
although such unstable solutions themselves are never realised in the flow dynamics obtained \textcolor{black}{through} direct
numerical simulations or experiments; see the review by Kawahara et al. (2012) and the references therein. 
This is exactly what can be expected from the famous Lorenz's toy model computation of atmospheric convection (Lorenz 1963); however the corresponding computations of the Navier-Stokes equations are not as easy.

In practice, unstable invariant solutions can be captured by the multi-dimensional Newton method, exactly \textcolor{black}{satisfying the complete} fluid dynamic equations within numerical accuracy.
Early years of the computations were motivated to explain \textcolor{black}{transition to turbulence} in \textcolor{black}{the} plane Couette flow.
\textcolor{black}{The laminar} plane Couette flow is always linearly stable. \textcolor{black}{However,} transition to turbulence occurs at moderate Reynolds numbers by \textcolor{black}{certain} perturbations characterised by streamwise roll-streak.
This apparent contradiction has been resolved by relating the nonlinear \textcolor{black}{finite amplitude} three-dimensional solutions and the subcritical transition using the dynamical systems theory.
Finding solutions in \textcolor{black}{a} linearly stable system is tricky, as we cannot use the bifurcation analysis from a linear critical point to find the first solution. 
\textcolor{black}{One possible resolution is to} consider the continuous (homotopic) deformation from the augmented system, where a nonlinear solution is known \textcolor{black}{for} the target system, wishing that the solution branch could be continued throughout the deformation.
The challenge to \textcolor{black}{determining} nontrivial solutions in \textcolor{black}{the} plane Couette flow began in Nagata's PhD project with his supervisor Busse, who proposed to use buoyancy as a homotopy parameter.
After \textcolor{black}{several} unsuccessful attempts, the first nonlinear solutions were found \textcolor{black}{by} Nagata (1990) where \textcolor{black}{the} Colioris force \textcolor{black}{was used} as a homotopy parameter, while their original approach was accomplished in Clever \& Busse (1992); \textcolor{black}{nowadays, these are known as} the Nagata-Busse-Clever (NBC) solutions. 
Subsequently the existence of \textcolor{black}{a} myriad of solutions in the flow has been uncovered; see Schmiegel (1999), Waleffe (2003), Itano \& Generalis (2009), Gibson et al. (2009). 
As remarked earlier, the importance of those solutions and some nearby dynamics in fully developed turbulence \textcolor{black}{were repeatedly} pointed out by Kawahara \& Kida (2001), Gibson et al. (2008), van Veen \& Kawahara (2011), Kreilos \& Eckhardt (2012) \textcolor{black}{and} Lustro et al. (2019).
\textcolor{black}{It should also be remarked that} in bistable systems such as plane Couette flow, some solutions play a \textcolor{black}{gatekeeping} role at the edge of the laminar and turbulent attractors; see Itano \& Toh (2001), Skufca, Yorke \& Eckhardt (2006), Wang, Gibson \& Waleffe (2007), Deguchi \& Hall (2016).

\textcolor{black}{Moreover, the numerical studies} unveiled the physical interpretation of the sustainment mechanism of roll-streak, called \textcolor{black}{the} self-sustaining process  (Waleffe 1997, Wang, Gibson \& Waleffe 2007). 
Before the numerical computation of the three-dimensional solutions, the appearance of streamwise roll in plane Couette flow was rather puzzling, because it is easy to show \textcolor{black}{through} energy balance that \textcolor{black}{the} two-dimensional streamwise independent roll cannot \textcolor{black}{sustain itself}. 
\textcolor{black}{Thus,} to support the roll component, some three dimensionarity of the flow is necessary.
The self-sustaining process explains the origin of the three-dimensional component by the wave-like \textcolor{black}{marginal} instability of the streak, which is the mean flow modulated by the roll via the lift-up mechanism.

There is another similar \textcolor{black}{yet} more mathematical research stream, the vortex/wave interaction theory (Hall \& Smith 1991), originally formulated for boundary layer flows rather than \textcolor{black}{the} plane Couette flow. 
While both the vortex/wave interaction and self-sustaining process theories concern the Reynolds number scaling of streamwise vortices when that number is large, there had been nonetheless an important difference.
Formally, the wave stability problem is inviscid when the large Reynolds number limit is taken, \textcolor{black}{producing} a singularity at the critical layer, where the wave speed coincides with the streak speed.
The vortex/wave interaction approach \textcolor{black}{uses} the method of matched asymptotic expansion to analyse the flow around the singularity. 
\textcolor{black}{The singular amplification of the wave amplitude therein is important, since it} causes the major feedback effect to the roll.
\textcolor{black}{Alternately,} the self-sustaining process retains the viscous effect in the wave equations. \textcolor{black}{Thus} the importance of the critical layer \textcolor{black}{was not noticed until the numerical discovery of the critical layer structure in invariant solutions by Wang et al. (2007). }

Motivated by \textcolor{black}{this} work, Hall \& Sherwin (2010) \textcolor{black}{applied the vortex/wave interaction theory to plane Couette flow. 
The extrapolated vortex/wave interaction result from infinite Reynolds numbers reproduced the Navier-Stokes result by Wang et al. (2007) remarkably well,} even at moderate Reynolds numbers. 
\textcolor{black}{To} solve the singular \textcolor{black}{asymptotic} problem, Hall \& Sherwin (2010) used \textcolor{black}{a} certain technical regularisation method. 
Subsequently, much simpler regularisation based on the `fictitious viscosity' \textcolor{black}{was} adopted (Blackburn et al. 2013; Deguchi \& Hall 2014b, 2016).
That hybrid approach, carried forward by the earlier work \textcolor{black}{of} Hall \& Horseman (1991), led \textcolor{black}{to the} identical formulation to the self-sustaining process as a result. \textcolor{black}{However,} an important point to note is that now, that formulation is backed up by fully rational asymptotic analysis \textcolor{black}{by} Hall \& Smith (1991).

Asymptotic developments of invariant solutions are \textcolor{black}{potentially} of enormous importance 
\textcolor{black}{since} the scaling of the flow dynamics in terms of the Reynolds number \textcolor{black}{is} the central interest in fluid dynamic \textcolor{black}{studies}.
The innovative combination of two mathematical tools, matched asymptotic analysis and unstable invariant solutions, has been employed in  
Deguchi et al. (2013), Deguchi \& Walton (2013a, 2013b, 2018), Deguchi \& Hall (2014a, 2014b, 2014c,  2015), Deguchi (2015, 2017), Dempsey et al. (2016), Ozcakir et al. (2016), sparked by excellent agreement \textcolor{black}{as} seen in Hall \& Sherwin (2010).
The advantage \textcolor{black}{of} the dual approach is that it is particularly useful for confirming or finding new asymptotic theories, \textcolor{black}{as} the simple structure of unstable invariant solutions enable a clean quantitative comparison of the theories with \textcolor{black}{complete} numerical results.
Moreover, some invariant solutions can be found at very high Reynolds numbers to produce remarkably accurate Reynolds number asymptotic scaling. For example, in a channel flow, the nonlinear solutions computed in Dempsey et al. (2016) \textcolor{black}{reached a} Reynolds number of order $10^8$, which is much higher than the maximum Reynolds number available by direct numerical simulations or experiments.
\textcolor{black}{Although the asymptotic solutions found so far have simpler structures compared with turbulence in direct numerical simulations, they provide important clues to understand the fundamental vortex structures without hand-wavy explanations.}

In this paper, we shall extend \textcolor{black}{the aforementioned} purely hydrodynamic studies to MHD flows. 
So far, invariant solutions in MHD equations have been studied in the flow under Kepler rotation, \textcolor{black}{which is linearly stable, in the absence of an external magnetic field.}
There \textcolor{black}{exist} much astrophysical interests in that flow \textcolor{black}{since} it may model flows around certain celestial bodies. 
Balbus \& Hawley (1991) introduced an external magnetic field to bring a linear instability mechanism, known as magneto-rotational instability, in the flow. However, the unknown precise origin of that external magnetic \textcolor{black}{field} at the beginning of the astrophysical flow formation posed the unanswered question of whether the subcritical transition \textcolor{black}{played} an important role in the instability mechanism. 

\textcolor{black}{Towards this end, the first invariant nonlinear dynamo solution without any external magnetic field} was found in \textcolor{black}{the} MHD rotating plane Couette flow by Rincon et al. (2007), and their results were subsequently extended to \textcolor{black}{the} rotating shearing box; see Riols et al. (2013).
\textcolor{black}{Note, these subcritical dynamos are fundamentally different from dynamos driven by centrifugal instability in cylindrical or spherical Couette flow apparatuses (Willis \& Barenghi 2002a; Nore et al. 2012; Marcotte \& Gissinger 2016). 
More recently, Guseva et al. (2017) succeeded in generating dynamos in a quasi-Keplerian cylindrical Couette flow, motivated by the rotating plane Couette flow and shearing box studies.}

The dynamo solutions found by Rincon et al. (2007) \textcolor{black}{and} Riols et al. (2013) are three-dimensional consistent to Cowling's anti-dynamo theorem (Cowling 1934), which states that there \textcolor{black}{are} no streamwise independent two-dimensional dynamo states. 
Inspired by the self-sustaining process, the origin of the three-dimensionality was explained by the magneto-rotational instability triggered by the induced magnetic streak (Rincon et al. 2008).
However, \textcolor{black}{these solutions were} found to be very difficult to continue large Reynolds number regime of astrophysical importance, opposed to the purely hydrodynamic plane Couette flow counterparts. 
\textcolor{black}{The problem is not limited to these nonlinear invariant solutions; whether dynamo turbulence as a whole can itself survive at large Reynolds numbers is still under debate (Walker et al. 2016; Walker \& Boldyrev 2017). }
In Deguchi \textcolor{black}{(2019)}, the straightforward extension of the vortex/wave interaction theory to MHD flows \textcolor{black}{was shown to be} possible, but the asymptotic state should be destroyed when a strong rotating effect presents.
Therefore, here, we only focus on the dynamo states driven in \textcolor{black}{the} non-rotating shear flows, \textcolor{black}{since} we are interested in \textcolor{black}{the} high Reynolds number regime. 

\textcolor{black}{
The kinematic-magnetic energy conversion mechanism through a shear is common in astrophysics and solar physics.
There has been much activity in the last ten years on what the dynamo community now refers to as the `shear dynamo' model (see e.g. Yousef et al. (2008), Heinemann et al. (2011), Tobias \& Cattaneo (2013), Teed \& Proctor (2017) and many other references in these articles).
However, the large-scale dynamo action seen in the non-rotating shear flows studied in this community can only be generated with the aid of  small-scale isotropic, non-helical forcing, whose origin is unexplained by the governing equations.
Till date, the numerical computation of self-consistent shear driven dynamos without small-scale forcing have received surprisingly less attention (see recent numerical study by Neumann \& Blackman 2017 and the references therein). }

The key idea in Deguchi \textcolor{black}{(2019)} is \textcolor{black}{the use of} the Alfv\'en wave instability as a driving mechanism of the roll-streak field, now defined for both velocity and magnetic fields. 
The Alfv\'en wave is stimulated by inhomogeneous hydrodynamic and magnetic streaky fields, and possesses the singularity \textcolor{black}{at} Alfv\'en resonant point; see Sakurai et al. (1991), Goossens et al. (1992) for \textcolor{black}{examples}. 
Near the resonant point, the wave is amplified to produce strong vortex and current sheets, which in turn produce a feedback effect to the hydrodynamic and magnetic roll-streak. 
\textcolor{black}{An important caveat in the asymptotic analysis of Deguchi (2019) is that the vortex/Alfv\'en wave interaction cycle is not realised without a some weak externally applied magnetic field. Nevertheless, the dynamo states are possible, as we shall soon see in this paper.}

After formulating our problem in the next section, \textcolor{black}{of selecting the} MHD plane Couette flow as a canonical flow configuration, we \textcolor{black}{will} begin our computation in section 3. 
The excitation of \textcolor{black}{the} three-dimensional magnetic field can be found by applying a uniform unidirectional magnetic or current field to the NBC solutions. 
\textcolor{black}{Particularly}, the small uniform spanwise magnetic field is \textcolor{black}{a most appropriate choice} to drive the vortex/Alfv\'en interaction states.
The invariant solutions weakly forced in this way \textcolor{black}{produce a} much stronger streamwise magnetic field, consistent to the theoretical analysis in \textcolor{black}{Deguchi (2019).}
In section 4, we \textcolor{black}{will} use an external uniform spanwise current field as a homotopy parameter to show that we can \textcolor{black}{continue the invariant MHD solution branch} even at the zero external magnetic field limit.
Motivated by this rather surprising result, another asymptotic theory \textcolor{black}{has been} formulated in Deguchi \textcolor{black}{(2019)} to explain the presence of \textcolor{black}{the} self-sustained shear driven dynamos, S$^3$ dynamos \textcolor{black}{in} short.
In the same section, the corresponding asymptotic problem is numerically solved using the hybrid approach. 
Finally, in section 5, we \textcolor{black}{presented our} conclusion.

\section{Computational method}

Consider electricaly conducting fluid flow between the perfectly insulating walls of \textcolor{black}{the} infinite extent placed at $y=\pm 1$. For $x,z$ directions, we assume \textcolor{black}{the} periodicity of the flow with the wavenumbers $\alpha,\beta$, respectively. 
Within that computational box, we numerically solve \textcolor{black}{for the} incompressible viscous resistive MHD equations, non-dimensionalised similar \textcolor{black}{to} Deguchi \textcolor{black}{(2019)}
\begin{subequations}\label{MHD}
\begin{eqnarray}
\frac{D\mathbf{v}}{Dt}
- (\mathbf{b}\cdot \nabla)\mathbf{b}
&=&-\nabla q + \frac{1}{R} \nabla^2 \mathbf{v} ,\\
\frac{D\mathbf{b}}{Dt}-(\mathbf{b}\cdot \nabla)\mathbf{v}
&=&\frac{1}{R_m} \nabla^2 \mathbf{b},\\
 \nabla \cdot  \mathbf{v}&=&0,\\
 \nabla \cdot \mathbf{b}&=&0,
\end{eqnarray}
\end{subequations}
where $\nabla=(\partial_{x},\partial_{y},\partial_{z})$, $D/Dt=(\partial_t+\mathbf{v}\cdot \nabla)$ and $q$ is the total pressure. 
\textcolor{black}{We denote the component form of the velocity and magnetic vectors as $\mathbf{v}=[u,v,w], \mathbf{b}=[a,b,c]$.}
The flow is driven by the given base flow $(u,v,w,a,b,c)=(U_b,0,W_b,A_b,0,C_b)$, which represents the laminar flow solution of the system and is \textcolor{black}{a} function of $y$. 
More specifically, we consider \textcolor{black}{the} plane Couette flow $U_b=y, W_b=0$, forced by some external magnetic field.
In \textcolor{black}{this case,} the hydrodynamic Reynolds number $R$ and the magnetic Reynolds number $R_m$ are defined using the half channel height and the wall speed as the length and velocity scales, respectively. \textcolor{black}{The ratio $P_m=R_m/R$ is the magnetic Prandtl number.}

The perturbation to the base flow $(\mathbf{U},\mathbf{B})=(\mathbf{u},\mathbf{b})-(\mathbf{U}_b,\mathbf{B}_b)$ must satisfy the no-slip and insulating conditions on the walls. 
\textcolor{black}{To} satisfy the latter conditions, the magnetic perturbation must match the magnetic field outside the computational domain. There should be no current in the outer region, \textcolor{black}{such that} the outer magnetic field must have a potential $\varphi$, \textcolor{black}{where} $\mathbf{b}=\nabla \varphi$. Hence, from the solenoidality, we must solve the Laplace equation for the potential $\triangle \varphi=0$ for $|y|>1$, requiting that the potential decays in the far-field $|y|\gg 1$. 

For simplicity, here, we only consider \textcolor{black}{the} perturbations of travelling wave form with the streamwise and spanwise wave speeds $s$ \textcolor{black}{and} $s_z$, respectively. 
It is convenient to use the coordinate attached to the travelling wave so that we have a steady problem.
In the transformed coordinate the operator $\partial_t$ in (\ref{MHD}) must be replaced by $-s\partial_x-s_z \partial_z$; note this means that \textcolor{black}{herein} and hereafter the $x,z$ coordinates are redefined.

The divergence free fields in the periodic box can be written by the toroidal-poloidal potentials as \textcolor{black}{follows:}
\begin{eqnarray}
\left[ \begin{array}{c} u\\ v\\ w \end{array} \right]=\left[ \begin{array}{c} \overline{\overline{u}}+\phi_{xy}+\psi_z\\ -\phi_{xx}-\phi_{zz} \\ \overline{\overline{w}}+\phi_{yz}-\psi_x \end{array} \right],\qquad
\left[ \begin{array}{c} a\\ b\\ c \end{array} \right]=\left[ \begin{array}{c} \overline{\overline{a}}+f_{xy}+g_z\\ -f_{xx}-f_{zz} \\ \overline{\overline{c}}+f_{yz}-g_x \end{array} \right].
\end{eqnarray}
Here the double overline represents the $x-z$ average
\begin{eqnarray}
\overline{\overline{(~)}}=\frac{\alpha \beta}{4\pi^2}\int_0^{2\pi/\alpha} \int_0^{2\pi/\beta} (~)dxdz.
\end{eqnarray}
Thus, the components $\overline{\overline{u}},\overline{\overline{w}},\overline{\overline{a}},\overline{\overline{c}}$ are functions of \textcolor{black}{only} $y$ and correspond to the mean components, which can further be decomposed into the base flow  and the mean perturbation \textcolor{black}{as the following:}
\begin{subequations}
\begin{eqnarray}
\overline{\overline{u}}(y)=U_b(y)+\overline{\overline{U}}(y),\qquad
\overline{\overline{w}}(y)=W_b(y)+\overline{\overline{W}}(y),\\
\overline{\overline{a}}(y)=A_b(y)+\overline{\overline{A}}(y),\qquad
\overline{\overline{c}}(y)=C_b(y)+\overline{\overline{C}}(y).
\end{eqnarray}
\end{subequations}

The equations for the potentials and the mean flows can be obtained from the \textcolor{black}{fluctuating} parts of the equations $\mathbf{e}_y \cdot \nabla \times \nabla \times (2.1a)$, $\mathbf{e}_y \cdot \nabla \times  (2.1a)$, $\mathbf{e}_y \cdot  (2.1b)$, $\mathbf{e}_y \cdot \nabla \times  (2.1b)$, and the mean parts of the equations $\mathbf{e}_x \cdot \overline{\overline{(2.1a)}}$, $\mathbf{e}_z \cdot \overline{\overline{(2.1a)}}$, $\mathbf{e}_x \cdot \overline{\overline{(2.1b)}}$, $\mathbf{e}_z \cdot \overline{\overline{(2.1b)}}$.
Now, we substitute the Fourier expansions with the basis $E_{m,n}=\exp[im\alpha x+in\beta z]$
\begin{subequations}
\begin{eqnarray}
\phi(x,y,z)=\sum_{m_0,n_0} E_{m_0,n_0}\widehat{\phi}_{m_0,n_0}(y),\qquad
\psi(x,y,z)=\sum_{m_0,n_0} E_{m_0,n_0}\widehat{\psi}_{m_0,n_0}(y),\\
f(x,y,z)=\sum_{m_0,n_0} E_{m_0,n_0}\widehat{f}_{m_0,n_0}(y),\qquad
g(x,y,z)=\sum_{m_0,n_0} E_{m_0,n_0}\widehat{g}_{m_0,n_0}(y),
\end{eqnarray}
\end{subequations}
(note, there is no mean component $(m_0,n_0)=(0,0)$ by definition) to those equations and operate
\begin{eqnarray}
\overline{\overline{E_{m,n}^{-1}(~)}}
\end{eqnarray}
to discretise the equations in $x$ and $z$. 
For the fluctuating parts the discretised equations are obtained as \textcolor{black}{follows:}
\begin{subequations}\label{flucequations}
\begin{eqnarray}
0&=& R^{-1}[ \widehat{\phi}''''-2L\widehat{\phi}''+L^2\widehat{\phi}]- i[m\alpha (\overline{\overline{u}}-s)+n\beta (\overline{\overline{w}}-s_z) ][\widehat{\phi}''-L\widehat{\phi} ]+i [m\alpha \overline{\overline{a}}+n\beta \overline{\overline{c}}][ \widehat{f}''-L\widehat{f} ]\nonumber
\\
 &&+i[m\alpha \overline{\overline{u}}''+n\beta \overline{\overline{w}}'']\widehat{\phi}-i[m\alpha \overline{\overline{a}}''+n\beta \overline{\overline{c}}'']\widehat{f}
+L^{-1}\sum_{m_1,n_1}\mathcal{N}^{(1)}(m,n,m_1,n_1),~~~~ \\ 
0&=&R^{-1}[\widehat{\psi}''-L\widehat{\psi}]-i[m\alpha (\overline{\overline{u}}-s)+n\beta (\overline{\overline{w}}-s_z) ]\widehat{\psi}+i[m\alpha \overline{\overline{a}}+n\beta \overline{\overline{c}}]\widehat{g} \nonumber \\
&& -i[m\alpha \overline{\overline{w}}'-n\beta \overline{\overline{u}}']\widehat{\phi}+i[m\alpha \overline{\overline{c}}'-n\beta \overline{\overline{a}}']\widehat{f} 
+L^{-1}\sum_{m_1,n_1}\mathcal{N}^{(2)}(m,n,m_1,n_1),~~~~\\
0&=&R_m^{-1} [\widehat{f}''-L\widehat{f}]-i[m\alpha (\overline{\overline{u}}-s) +n\beta (\overline{\overline{w}}-s_z)   ]\widehat{f} +i[m\alpha \overline{\overline{a}}+n\beta \overline{\overline{c}}]\widehat{\phi} \nonumber \\
&&+L^{-1}\sum_{m_1,n_1}\mathcal{N}^{(3)}(m,n,m_1,n_1),~~~~\\
0&=&R_m^{-1}[\widehat{g}''-L\widehat{g}]-i[m\alpha (\overline{\overline{u}}-s)+n\beta (\overline{\overline{w}}-s_z) ]\widehat{g}+i[m\alpha \overline{\overline{a}}+n\beta \overline{\overline{c}}]\widehat{\psi} \nonumber \\ 
&& -i[m\alpha \overline{\overline{c}}'-n\beta \overline{\overline{a}}']\widehat{\phi}+i[m\alpha \overline{\overline{w}}'-n\beta \overline{\overline{u}}']\widehat{f} 
+L^{-1}\sum_{m_1,n_1}\mathcal{N}^{(4)}(m,n,m_1,n_1),~~~~
\end{eqnarray}
\end{subequations}
where the summations are taken for $m_2=m-m_1$, $n_2=n-n_1$ with the nonlinear terms shown in Appendix A.
Here, we have used the shorthand \textcolor{black}{notation $L=(m\alpha)^2+(n\beta)^2$}
and the Fourier transformed variables are abbreviated as \textcolor{black}{below:}
\begin{eqnarray}
\widehat{\phi}=\widehat{\phi}_{m,n}(y),\qquad \widehat{\phi}_1=\widehat{\phi}_{m_1,n_1}(y),\qquad \widehat{\phi}_2=\widehat{\phi}_{m_2,n_2}(y), ~~~~\text{etc.}
\end{eqnarray}
Likewise, the mean part of the Fourier transformed equations are obtained as \textcolor{black}{
\begin{subequations}\label{meanequations}
\begin{eqnarray}
\overline{\overline{U}}''+i\sum_{m_1,n_1}\mathcal{N}_0^{(1)}(m_1,n_1)=0,\qquad
\overline{\overline{W}}''+i\sum_{m_1,n_1}\mathcal{N}_0^{(2)}(m_1,n_1)=0,\\
\overline{\overline{A}}''+i\sum_{m_1,n_1}\mathcal{N}_0^{(3)}(m_1,n_1)=0, \qquad
\overline{\overline{C}}''+i\sum_{m_1,n_1}\mathcal{N}_0^{(4)}(m_1,n_1)=0,
\end{eqnarray}
\end{subequations}
where, the summations are taken for $m_1=-m_2$, $n_1=-n_2$; the nonlinear terms are again shown in Appendix A.}

\textcolor{black}{To} further discretise the equations in $y$,
we expand the Fourier coefficients by modified Chebyshev polynomials. 
The no-slip conditions on the walls
\begin{eqnarray}
\widehat{\phi}=\widehat{\phi}'=\widehat{\psi}=\overline{\overline{u}}=\overline{\overline{w}}=0 \qquad \text{at}\qquad y=\pm 1
\end{eqnarray}
are satisfied using the basis functions
\begin{eqnarray*}
\widehat{\phi}_{m,n}(y)=\sum_l X_{l,m,n}^{(1)}(1-y^2)^2T_l(y),\qquad
\widehat{\psi}_{m,n}(y)=\sum_l X_{l,m,n}^{(2)}(1-y^2)T_l(y),\\
\overline{\overline{u}}(y)=\sum_l X_{l,0,0}^{(1)}(1-y^2)T_l(y),\qquad
\overline{\overline{w}}(y)=\sum_l X_{l,0,0}^{(2)}(1-y^2)T_l(y).\\
\end{eqnarray*}
In order to find the appropriate basis for the magnetic part, we need some analysis of the Fourier transformed outer magnetic potential
\begin{eqnarray}
\varphi(x,y,z)=\sum_{m_0,n_0} E_{m_0,n_0}\widehat{\varphi}_{m_0,n_0}(y).
\end{eqnarray}
The potential satisfying the Laplace equation with the required far-field conditions can be found as
\begin{eqnarray}
\widehat{\varphi}_{m_0,n_0}=\left \{
\begin{array}{c}
\Phi_{m_0,n_0} e^{-\sqrt{L}y}\qquad \text{if}\qquad y>1,\\
\Psi_{m_0,n_0} e^{\sqrt{L}y}\qquad \text{if}\qquad y<-1,
\end{array}
\right .
\end{eqnarray}
where, $\Phi_{m_0,n_0},\Psi_{m_0,n_0}$ are constants and $\Phi_{0,0}=\Psi_{0,0}=0$.
As mentioned above, the magnetic perturbation of the flow must match this outer field on the walls. Thus, the mean part should vanish and the fluctuating parts \textcolor{black}{should} satisfy 
\begin{subequations}
\begin{eqnarray}
im\alpha \widehat{f}'+in\beta \widehat{g} =im\alpha \widehat{\varphi},\\
L\widehat{f}=\mp \sqrt{L}\widehat{\varphi},\\
in\beta \widehat{f}'-im\alpha \widehat{g} =in\beta \widehat{\varphi},
\end{eqnarray}
\end{subequations}
at $y=\pm 1$.
Therefore, we finally \textcolor{black}{obtain} the boundary conditions
\begin{eqnarray}
\widehat{f}'\pm\sqrt{L}\widehat{f}=\widehat{g}=\overline{\overline{a}}=\overline{\overline{c}}=0 \qquad \text{at}\qquad y=\pm 1,\label{insulating}
\end{eqnarray}
which is exactly the narrow-gap limit version of the insulating conditions used for Taylor-Couette flow; see Roberts (1964), Willis \& Barenghi (2002b), Rudiger et al. (2003). The choice of the basis functions \textcolor{black}{
\begin{eqnarray*}
\widehat{f}_{m,n}(y)&=&\sum_l X_{l,m,n}^{(3)}\left \{(1-y^2)T_l(y)+\frac{1+(-1)^l}{\sqrt{L}}+\frac{1-(-1)^l}{\sqrt{L}+1} y \right \},\\
\widehat{g}_{m,n}(y)&=&\sum_l X_{l,m,n}^{(4)}(1-y^2)T_l(y),\\
\overline{\overline{a}}(y)&=&\sum_l X_{l,0,0}^{(3)}(1-y^2)T_l(y),\qquad
\overline{\overline{c}}(y)=\sum_l X_{l,0,0}^{(4)}(1-y^2)T_l(y),
\end{eqnarray*}
ensure (\ref{insulating}).}

For the numerical purpose, we truncate the expansion and only use the coefficients $l\in[0,K]$, $m_0\in [-M,M]$, $n_0\in [-N,N]$. 
Evaluating (\ref{flucequations}) and (\ref{meanequations}) at the collocation points
\begin{eqnarray}
y_k=\cos \left (\frac{k+1}{K+2}\pi \right ),\qquad k=0,\dots,K,
\end{eqnarray}
the resultant equations with the Fourier discretisation parameters $|m|<M$ and $|n|<N$ constitute the algebraic equations for the spectral coefficients $X^{(j)}_{l,m,n}$ and the phase speeds $s,s_z$. 
Note, \textcolor{black}{approximately} half of the coefficients/equations are redundant \textcolor{black}{due to} the reality conditions $\Re (X^{(j)}_{l,m,n})=\Re (X^{(j)}_{l,-m,-n}), \Im (X^{(j)}_{l,m,n})=-\Im (X^{(j)}_{l,-m,-n})$. 
\textcolor{black}{Moreover,} the two extra equations needed for the phase speeds are some conditions that can eliminate the freedom of the solution associated with the arbitrary shift in $x$ and $z$. 
The algebraic equations are quadratic so the exact expression of the Jacobean matrix can be computed straightforwardly. 
The range of \textcolor{black}{the} numerical resolutions used are $K\in[80,120], M\in[3,10], N\in[24,34]$.

\section{\textcolor{black}{Nonlinear MHD states} with external magnetic fields}

We \textcolor{black}{used the} plane Couette flow $U_b=y,W_b=0$ with $(\alpha,\beta,P_m)=(1,2,1)$ throughout the computations in this paper. 
In the absence of any magnetic \textcolor{black}{field}, the MHD equations become the Navier-Stokes equations. Therefore, the hydrodynamic solutions obtained in the previous studies constitute the solutions of the MHD plane Couette flow \textcolor{black}{in the absence of an} external magnetic field. 
They serve a good starting point of the Newton continuation to find the \textcolor{black}{MHD} solution branch, as the imposed external magnetic fields $A_b,C_b$ stimulate \textcolor{black}{the} three-dimensional magnetic fields.

\subsection{Symmetry of the MHD solutions}

The symmetries of the dynamo solutions are important, because we fully use them to reduce computational costs.
Here, we shall clarify how the form of the external magnetic fields affects the symmetry of the solutions. 

We \textcolor{black}{have begun} the symmetry analysis \textcolor{black}{using} the purely hydrodynamic cases. 
The results to be obtained below are similar to \textcolor{black}{that of} Gibson et al. (2009). 
\textcolor{black}{First, we have aimed} to derive all possible symmetry classes of the solutions, assuming the invariance of the equations under the operations
\begin{subequations}\label{symeq}
\begin{eqnarray}
\sigma_x[u,v,w](x,y,z)=[-u,v,w](-x,y,z),\\
\sigma_y[u,v,w](x,y,z)=[u,-v,w](x,-y,z),\\
\sigma_z[u,v,w](x,y,z)=[u,v,-w](x,y,-z),\\
\tau_x[u,v,w](x,y,z)=[u,v,w](x+\pi/\alpha,y,z),\\
\tau_z[u,v,w](x,y,z)=[u,v,w](x,y,z+\pi/\beta).
\end{eqnarray}
\end{subequations}
The above operations are motivated by the invariance of the Navier-Stokes equations under the reflection and shift in $x,y,$ or $z$ directions. Here, we \textcolor{black}{have not considered} the shift in \textcolor{black}{the} $y$ direction, because the fluid motion is limited by the walls. For $x,z$ directions, we only focus on \textcolor{black}{the} half-shift for the sake of simplicity (if we \textcolor{black}{allowed} general shifts, it \textcolor{black}{would} just produces a few special cases). We \textcolor{black}{have maintained the abbreviations used by} Gibson et al. (2009) for the combined operations, namely, $\sigma_{xy}=\sigma_x\sigma_y$, etc.

\textcolor{black}{To} classify the solutions in terms of their symmetry, we check whether they are invariant under the operations in (\ref{symeq}) and/or their combinations. However, in order to classify the solutions in terms of their symmetry, some operations should be excluded from the consideration from the reasons \textcolor{black}{listed} below.
\begin{enumerate}
\item It is desirable to choose the periodic box \textcolor{black}{such} that there is no identical flow copy within the box. 
\textcolor{black}{Thus,} any solution invariant under $\tau_x$ or $\tau_z$ must be excluded.

\item Whether a solution is invariant under a reflection operator actually depends on the choice of the origin. 
\textcolor{black}{To} develop the theory independent of that choice, hereafter, we \textcolor{black}{shall maintain} that a solution is invariant under an operation when we can choose at least one coordinates where this invariance appears. This \textcolor{black}{implies} that \textcolor{black}{if an} operation like $\tau_x\sigma_x$ can be omitted because the invariance, it \textcolor{black}{would be} equivalent to that of $\sigma_x$.

\item The operations not consistent with the base flow, should be excluded.
For example, if the flow is invariant under $\sigma_x$, this means that $\overline{\overline{u}}=0$ and, for the base part, $U_b-s=0$ (recall that we are now using the travelling wave coordinates). \textcolor{black}{Thus,} whenever $U_b-s\neq 0$, we \textcolor{black}{have to} exclude $\sigma_x$ from the possible operations. 
The relationship of the operations and the expected restriction for the base flow when the flow is invariant to the corresponding operator is summarised in table 1. (All magnetic effects in the table should be omitted at this stage. \textcolor{black}{Additionally,} the operator $\gamma$ is to be defined shortly for the MHD equations).
\end{enumerate}

Given the extended meaning of the invariance and the set of operations under consideration, we now denote the classes of the solutions by the curly bracket, within which the invariant operators are \textcolor{black}{to be} written. For \textcolor{black}{the} plane Couette flow, there are five possible operations \textcolor{black}{that} we must consider, $\sigma_z,\tau_x\sigma_z,\sigma_{xy},\tau_z\sigma_{xy},\sigma_{xyz}$, in view of (a)-(c) above and table 1. The NBC solution belongs to the class $\{\tau_x\sigma_z, \tau_z\sigma_{xy},\sigma_{xyz}\}$, \textcolor{black}{since} the solution is invariant under $\tau_x\sigma_z, \tau_z\sigma_{xy},\sigma_{xyz}$ and not invariant under $\sigma_z,\sigma_{xy}$. Note, the solutions in this class must be steady, \textcolor{black}{as} the shifted basic flow $U_b-s=y-s$ is an odd function, and $W_b-s_z=-s_z$ should \textcolor{black}{vanish}. 

All possible classes for \textcolor{black}{the} plane Couette flow solutions can be found fairly easily. The least symmetric class is of course $\{\}$, where the solutions \textcolor{black}{belonging} to this class do not have any symmetry. The next less symmetric five classes can be found straightforwardly: $\{\sigma_z\}$, $\{\tau_x\sigma_z\}$, $\{\sigma_{xy}\}$, $\{\tau_z\sigma_{xy}\}$, $\{\sigma_{xyz}\}$. 
If the solutions are invariant under two operations, they must be invariant under the combined operation of those two. Thus, the next six higher symmetric classes are made of three elements: $\{\sigma_z, \sigma_{xy},\sigma_{xyz}\}$, $\{\sigma_z, \tau_z\sigma_{xy},\sigma_{xyz}\}$, $\{\tau_x\sigma_z, \sigma_{xy},\sigma_{xyz}\}$, $\{\tau_x\sigma_z, \tau_z\sigma_{xy},\sigma_{xyz}\}$, $\{\sigma_{xy}, \tau_z\sigma_{xy},\tau_{xz}\}$, $\{\sigma_{z}, \tau_x\sigma_{z},\tau_{xz}\}$ (Note, in the latter class, the origin used for the operations $\sigma_{z}$, $\tau_x\sigma_{z}$ must differ by a half shift in the $z$ direction).
Furthermore, we can find one highest symmetric class of six elements $\{\sigma_z,\tau_x\sigma_z,\sigma_{xyz},\sigma_{xy},\tau_z\sigma_{xy},\tau_{xz}\}$. 

\begin{table}
  \begin{center}
    \begin{tabular}{ccccc}
     Operators & $U_b(y)-s$ & $W_b(y)-s_z$ & $A_b(y)$ & $C_b(y)$ \\
     $\sigma_x$ & zero & any & zero & any \\ 
     $\sigma_y$ & even & even & even & even \\ 
     $\sigma_z$ & any & zero & any & zero \\ 
     $\sigma_{yz}$ & even & odd & even & odd \\ 
     $\sigma_{xz}$ & zero & zero & zero & zero \\ 
     $\sigma_{xy}$ & odd & even & odd & even \\ 
     $\sigma_{xyz}$ & odd & odd & odd & odd \\ 
     $\gamma\sigma_x$ & zero & any & any & zero \\ 
     $\gamma\sigma_y$ & even & even & odd & odd \\ 
     $\gamma\sigma_z$ & any & zero & zero & any \\ 
     $\gamma\sigma_{yz}$ & even & odd & odd & even \\ 
     $\gamma\sigma_{xz}$ & zero & zero & any & any \\ 
     $\gamma\sigma_{xy}$ & odd & even & even & odd \\ 
     $\gamma\sigma_{xyz}$ & odd & odd & even & even \\ 
    \end{tabular}
  \end{center}
\caption{The required symmetry for the base flows when the solution is invariant under the corresponding operator.} 
\label{sample-table}
\end{table}

\begin{figure}
\centering
\includegraphics[scale=0.27]{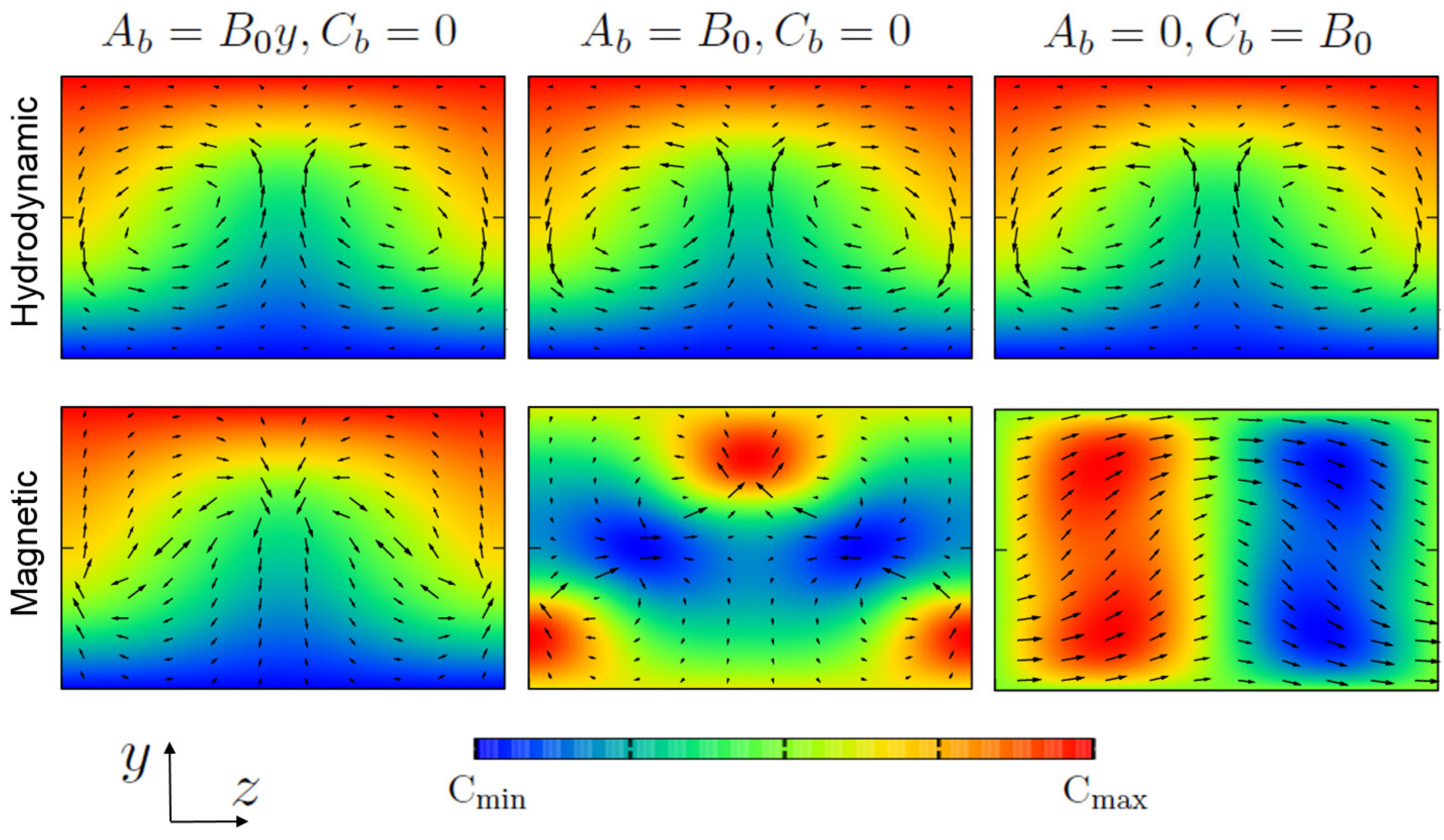} 
\caption{
The $x$-averaged field of the NBC solutions modified by the external magnetic fields. 
The left panels use $A_b=B_0y,C_b=0$ with $B_0=0.5$ (see figure 8), the middle panels use $A_b=B_0,C_b=0$ with $B_0=0.03$ (see figure 13), and the right panels use $A_b=0,C_b=B_0$ with $B_0=0.0002$ (see figure 3).
\textcolor{black}{The upper} panels are the hydrodynamic roll-streak ($(\text{C}_{\text{min}},\text{C}_{\text{max}})=(-1,1)$ for all three figures).
\textcolor{black}{The lower} panels are the magnetic roll-streak ($(\text{C}_{\text{min}},\text{C}_{\text{max}})=(-0.5,0.5), (0.013,0.04), (-0.016,0.016)$ from left to right). \textcolor{black}{These are all steady solutions of (\ref{MHD}).}
}
\label{fig:longwave}
\end{figure}

\begin{figure}
\centering
\includegraphics[scale=0.27]{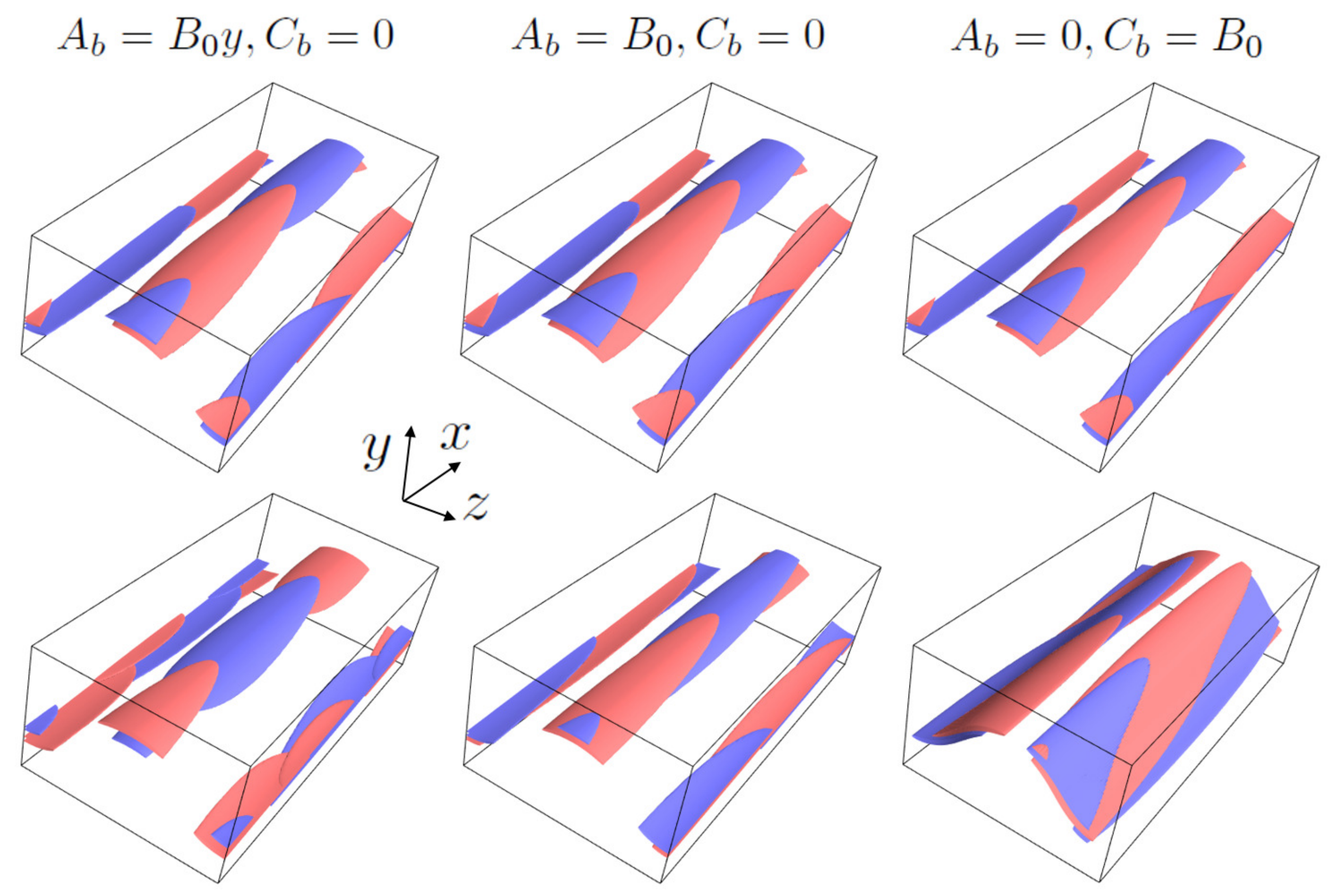} 
\caption{The same visualisation as figure 1 but for isosurface of 50\% streamwise vorticity (upper panels) and current (lower panels). Red: positive, blue: negative.}
\label{fig:longwave}
\end{figure}

\textcolor{black}{To} extend the above symmetry argument to the MHD flows, we must decide the action of the operators on the magnetic components. 
The most natural choice would be to transform the magnetic parts $a,b,c$ \textcolor{black}{in} the same manner as the hydrodynamic parts $u,v,w$. 
In fact, the MHD equations are invariant under the operations (\ref{symeq}) extended in this \textcolor{black}{manner.}
However, the MHD equations are also invariant under the flip of the polarity
\begin{eqnarray}
\gamma[u,v,w,a,b,c](x,y,z)=[u,v,w,-a,-b,-c](x,y,z)\label{symgamma}
\end{eqnarray}
which brings \textcolor{black}{further} complication. 
The above operator does not affect the hydrodynamic part.
\textcolor{black}{Thus, even if} all the extended classes given below preserve the symmetries of the NBC class for the hydrodynamic part, for the magnetic part we have completely different symmetry:
\begin{subequations}
\begin{eqnarray}
\{\tau_x\sigma_z, \tau_z\sigma_{xy},\sigma_{xyz}\},\label{sym1}\\
\{\gamma \tau_x\sigma_z, \tau_z\sigma_{xy},\gamma \sigma_{xyz}\},\label{sym2}\\
\{\tau_x\sigma_z, \gamma \tau_z\sigma_{xy},\gamma \sigma_{xyz}\},\label{sym3}\\
\{\gamma \tau_x\sigma_z, \gamma \tau_z\sigma_{xy},\sigma_{xyz}\}.\label{sym4}
\end{eqnarray}
\end{subequations}
\textcolor{black}{The symmetry of the mean flow inferred from the operators in the above classes are summarised in table 1.
These symmetries must be consistent with the applied base magnetic field $A_b,C_b$.}

For the first case (\ref{sym1}), from table 1 \textcolor{black}{we see} that $A_b(y)$ should be an odd function and $C_b(y)=0$. This is the expected result, \textcolor{black}{since} for this class, the hydrodynamic and magnetic fields must have identical symmetries.
The left panels in \textcolor{black}{figures} 1 and 2 show the flow field of this class computed by imposing the base magnetic field $A_b=B_0y, C_b=0,$ corresponding to the uniform spanwise current. 
Here the constant $B_0$ represents the intensity of the external field. 
Figure 1 shows the roll-streak field $\overline{\mathbf{v}}$ and $\overline{\mathbf{b}}$, defined by the streamwise average
\begin{eqnarray*}
\overline{(~)}=\frac{\alpha }{2\pi}\int_0^{2\pi/\alpha}  (~)dx.
\end{eqnarray*}
(Note, the definition of the overline is different from that in Deguchi \textcolor{black}{(2019)}.)
Figure 2 is the total field visualised by the isosurfaces of the streamwise vorticity and current. 
\textcolor{black}{Here,} the emergence of the strong vortex and current sheets can be seen, and they are in fact the signatures of the amplified Alfv\'en wave due to the resonance.

Table 1 suggests $C_b(y)=0$ for the second case (\ref{sym2}). \textcolor{black}{However,} now $A_b(y)$ should be an even function. 
Imposing the uniform streamwise magnetic field, $A_b=B_0, C_b=0$, we can generate the three dimensional magnetic field of \textcolor{black}{a} different symmetry, as shown in the middle panels of figures 1 \textcolor{black}{and} 2. 

In the third case (\ref{sym3}), we must turn off the streamwise magnetic field \textcolor{black}{such} that $A_b(y)=0$. \textcolor{black}{Instead, we} can apply a spanwise magnetic field of an even function $C_b(y)$. In the computation, we use the spanwise uniform magnetic field $A_b=0, C_b=B_0$. \textcolor{black}{The solution is} visualised in the right panels of \textcolor{black}{figures 1 and 2; the symmetry of it} is actually the same as that observed in the MHD rotating plane Couette flow computation by Rincon et al. (2007). 

\textcolor{black}{The fourth case (\ref{sym4})} can be generated by the spanwise magnetic field of \textcolor{black}{the} odd function. \textcolor{black}{However, we shall omit this case here.}

The external magnetic fields mentioned above are the only possible cases to preserve the NBC symmetry for the hydrodynamic field. In fact, it is easy to theoretically/numerically confirm that if, for example, the base magnetic field $A_b=B_0y, C_b=B_0$ is applied, the NBC symmetry seen in the hydrodynamic part is destroyed for $B_0\neq 0$.

\subsection{Asymptotic scaling of the vortex/Alfv\'en wave interaction states}

\begin{table}
  \begin{center}
    \begin{tabular}{cccccc}
    ~~~~~~~~~& Streak &  Roll & Outer wave & Inner wave & Wave amplitude \\
    Type 1 & $O(R^0)$ & $O(R^{-1})$ & $O(R^{-1})$ & $O(R^{-2/3})$ & $O(R^{-5/6})$ \\ 
    Type 2 & $O(R^0)$ & $O(R^{-1})$ & $O(R^{-7/6})$ & $O(R^{-5/6})$ & $O(R^{-1})$ 
    \end{tabular}
  \end{center}
\caption{The scaling of the vortex/Alfv\'en interaction states \textcolor{black}{obtained in the asymptotic theory by Deguchi (2019).} } 
\label{sample-table}
\end{table}
\textcolor{black}{This section aims} to check one of the nonlinear, three-dimensional \textcolor{black}{MHD} theories proposed in Deguchi \textcolor{black}{(2019)}, the vortex/Alfv\'en wave interaction, using the MHD solutions supported by the uniform spanwise magnetic field (corresponding to the third case (\ref{sym3})). 
The theory concerns the interaction of the roll-streak (i.e. $\overline{\mathbf{v}}$ and $\overline{\mathbf{b}}$) components and the wave components defined by $\mathbf{v}-\overline{\mathbf{v}}$ and $\mathbf{b}-\overline{\mathbf{b}}$. 
The \textcolor{black}{theoretical} leading order scaling of those components are summarised in table 2.
The $O(1)$ streak and $O(R^{-1})$ roll fields are typical in the vortex/wave interaction and \textcolor{black}{the} self-sustaining process theories.
\textcolor{black}{
The wave on top of the roll-streak has a finite amplitude but iis small enough to be originated from the linear marginal instability of the streak. This implies that only the wave neutral to the streak field has the leading order contribution; hence, the leading order flow can be written as
\begin{eqnarray}
\left[ \begin{array}{c} u\\ v\\ w\\a\\b\\c\\q \end{array} \right]=
\left[ \begin{array}{c} \overline{u}\\ \overline{v}\\ \overline{w}\\ \overline{a}\\ \overline{b}\\ \overline{c}\\  \overline{q}  \end{array} \right]
+\left \{ e^{i \alpha x}
\left[ \begin{array}{c} \widetilde{u}\\ \widetilde{v}\\ \widetilde{w}\\ \widetilde{a}\\ \widetilde{b}\\ \widetilde{c}\\\widetilde{q}\end{array} \right]
+\text{c.c.}\label{expincomp}
\right \},
\end{eqnarray}
where $\widetilde{u}, \widetilde{v}, \widetilde{w}, \widetilde{a}, \widetilde{b}, \widetilde{c}, \widetilde{q}$ are the complex functions of $y,z$, $\overline{u}=U_b+\overline{U}$, and c.c. stands for the complex conjugate.
Then, from (\ref{MHD}), we can show that the stability problem at the asymptotic limit is governed by the following equation:
\begin{eqnarray}
\left(\frac{\widetilde{q}_y}{(\overline{u}-s)^2-\overline{a}^2} \right )_y
+\left(\frac{\widetilde{q}_z}{(\overline{u}-s)^2-\overline{a}^2} \right )_z
-\alpha^2\frac{\widetilde{q}}{(\overline{u}-s)^2-\overline{a}^2}=0.
\end{eqnarray}
}

The singularity occurs in the inviscid wave problem whenever the resonant conditions $\overline{u}-s-\overline{a}=0$ or $\overline{u}-s+\overline{a}=0$ are satisfied. 
Near those curves, the wave is amplified singular \textcolor{black}{in a} manner and thus, the dissipative layers of thickness $O(R^{-1/3})$ surrounding the singularity \textcolor{black}{are} necessary to regularise the flow \textcolor{black}{therein}. 
In table 2, we \textcolor{black}{have showed} the theoretical wave scaling for the inside and outside of the dissipative layer.
As shown in Deguchi \textcolor{black}{(2019),} two types of the inner/outer wave scaling are possible, depending on the two resonant layers are well-separated (type 1) or degenerated (type 2).
\textcolor{black}{The type 2 interaction occurs when the distance between the two resonant curves is smaller than the dissipative layer thickness. In this case, two layers should merge and the singularity therein becomes stronger than that in the type 1 scenario, thereby changing the required wave amplitude to support the roll-streak field.}

\begin{figure}
\centering
\includegraphics[scale=1.1]{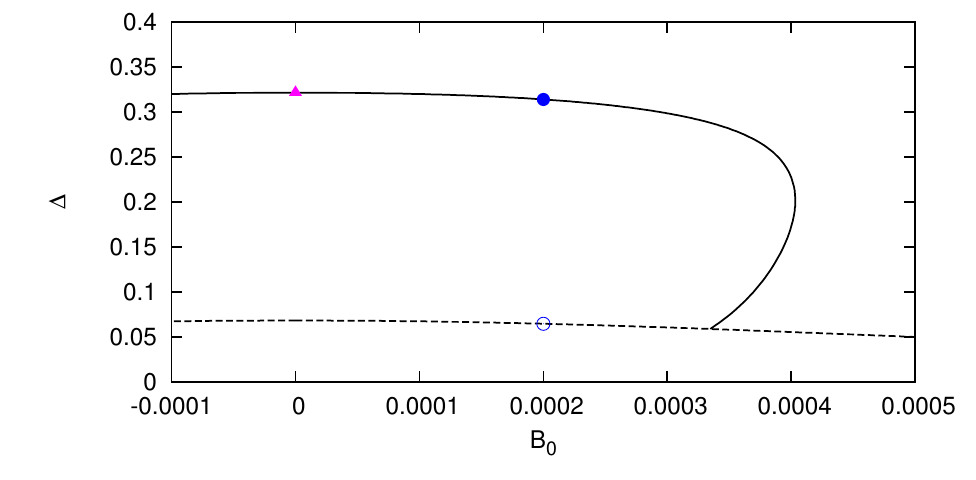} 
\caption{The bifurcation diagram for the external field $A_b=0, C_b=B_0$. The solid curve is the sinuous mode branch continued from the NBC solution (the magenta triangle).
The shear on the upper wall ($\Delta=\overline{\overline{u}}'(1)-1$) is used to measure the solution. 
The visualisations in the right panels of figures 1,2 correspond to the filled circle. 
The dashed curve is the mirror-symmetric mode branch. The open circle corresponds to the visualisation in figure 4.
}
\label{fig:longwave}
\end{figure}

\begin{figure}
\centering
\includegraphics[scale=0.46]{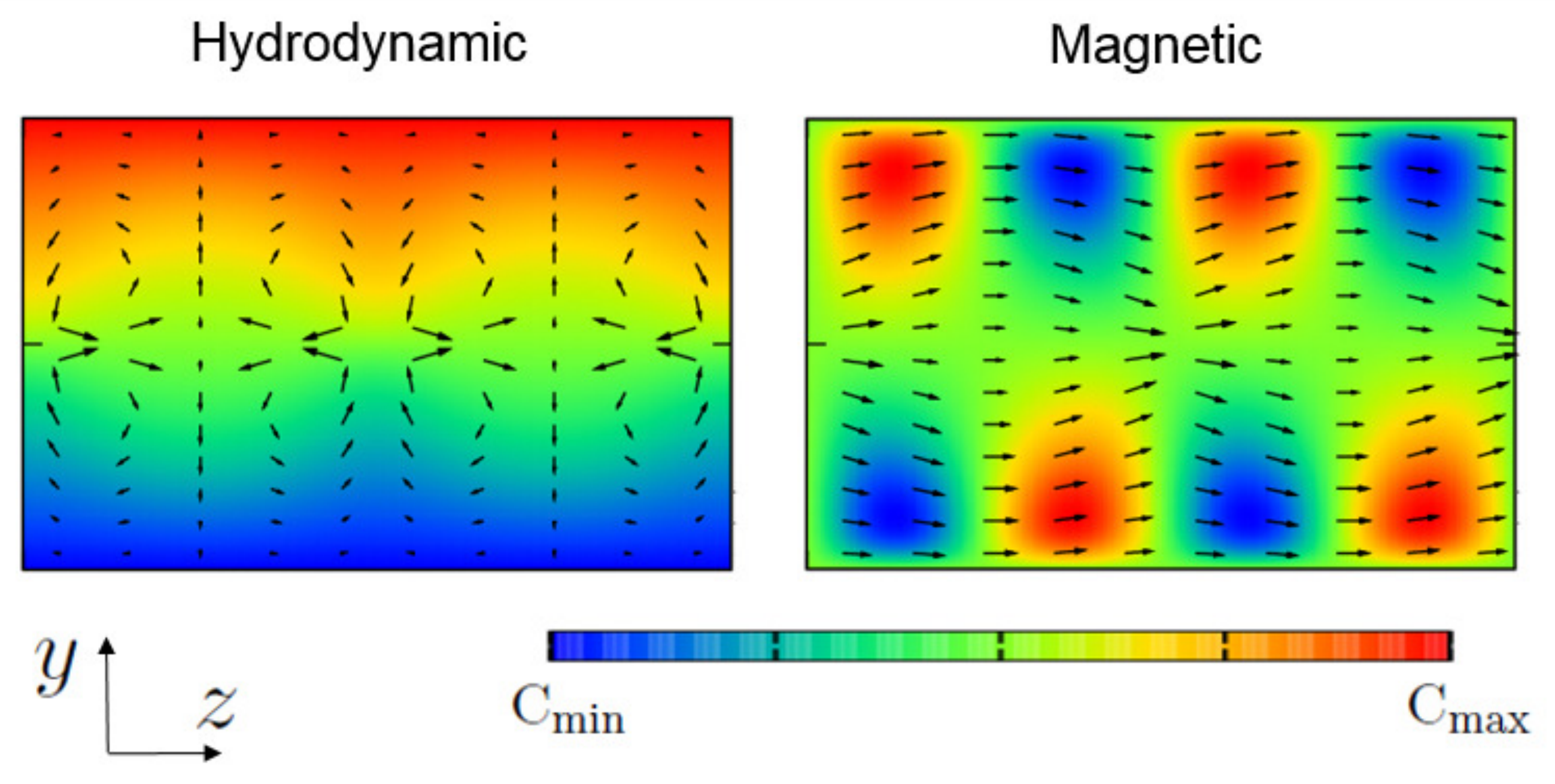} \\
\includegraphics[scale=0.46]{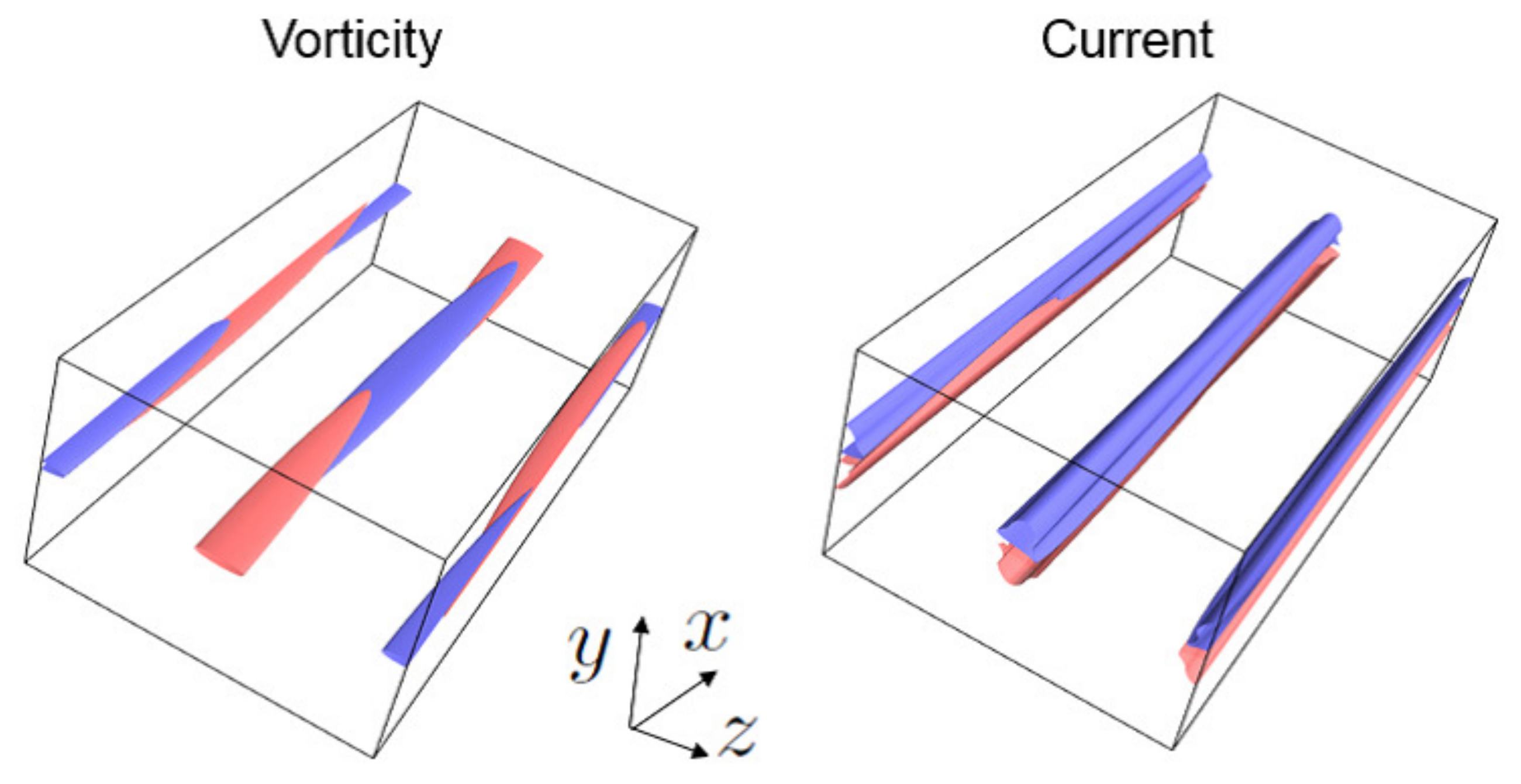} 
\caption{The same plot as figures 1 and 2 but for the mirror-symmetric mode computed in figure 3 ($B_0=0.0002$). The upper panels are the $x$-averaged fields (left panel: $(\text{C}_{\text{min}},\text{C}_{\text{max}})=(-1,1)$, right panel: $(\text{C}_{\text{min}},\text{C}_{\text{max}})=(-0.0016,0.0016)$). The lower panels are the 50 \% streamwise vorticity and current.}
\label{fig:longwave}
\end{figure}

For the vortex/Alfv\'en wave interaction, the leading order magnetic field cannot be maintained without some external magnetic field \textcolor{black}{owed to} the lack of energy input to the magnetic roll equations. 
Here, we apply the small uniform external magnetic field $A_b=0,C_b=B_0$ with $B_0\sim O(R^{-1})$ expecting \textcolor{black}{this} will stimulate \textcolor{black}{a} larger magnetic field in the flow through the \textcolor{black}{induction}.
Figure 3 shows the bifurcation diagram obtained by applying \textcolor{black}{the} base magnetic field to the NBC solutions (magenta triangle). 
The vertical axis of this diagram is the deviation of the shear on the upper wall, \textcolor{black}{$\Delta\equiv\overline{\overline{u}}'(1)-1$.}
The continued MHD solutions, hereafter called the sinuous mode, generate the three-dimensional magnetic field as seen in figures 1 \textcolor{black}{and} 2. 
Further increasing \textcolor{black}{the} $B_0$, the solution branch experiences a turning point at $B_0\approx0.0004$; on the way back, it connects to another solution branch shown by the dashed curve. 
The solutions on the latter branch, which we \textcolor{black}{shall} call the mirror-symmetric mode, belong to the highly symmetric class $\{\sigma_z,\tau_x\sigma_z,\gamma \sigma_{xyz},\gamma \sigma_{xy},\gamma \tau_z\sigma_{xy},\tau_{xz}\}$; see the visualisation shown in figure 4. 
At the hydrodynamic limit ($B_0=0$), the solution reduces to the known hydrodynamic solution found by Itano \& Generalis (2009) and Gibson et al. (2009) \textcolor{black}{(the solution is EQ7 in the latter paper, and \textcolor{black}{belonging} to the highest symmetry class $\{\sigma_z,\tau_x\sigma_z,\sigma_{xyz},\sigma_{xy},\tau_z\sigma_{xy},\tau_{xz}\}$ mentioned earlier).
The mirror-symmetric solution possesses a real eigenvalue, which crosses the imaginary axis at the bifurcation point in figure 3. This eigenmode breaks the mirror symmetry; hence the sinuous mode branch appears as a result of pitchfork bifurcation. Similar bifurcation has been found by Deguchi \& Hall (2014a) but for the purely hydrodynamic case.}

Figure 5 compares the vorticity of the sinuous and mirror-symmetric mode solutions at $x=0$. 
There are two resonant curves for the sinuous mode, so this is formally type 1, whilst for the mirror-symmetric modes the two resonant positions must be degenerated at $y=0$ by their symmetry \textcolor{black}{and thus} could be regarded as a good example of type 2.
The vorticity is dominated by the amplified wave component near the resonant curves, consistent to the theory.
Similar amplification of the magnetic wave component can be found by visualising the streamwise current; see figure 2.

\begin{figure}
\centering
\includegraphics[scale=0.35]{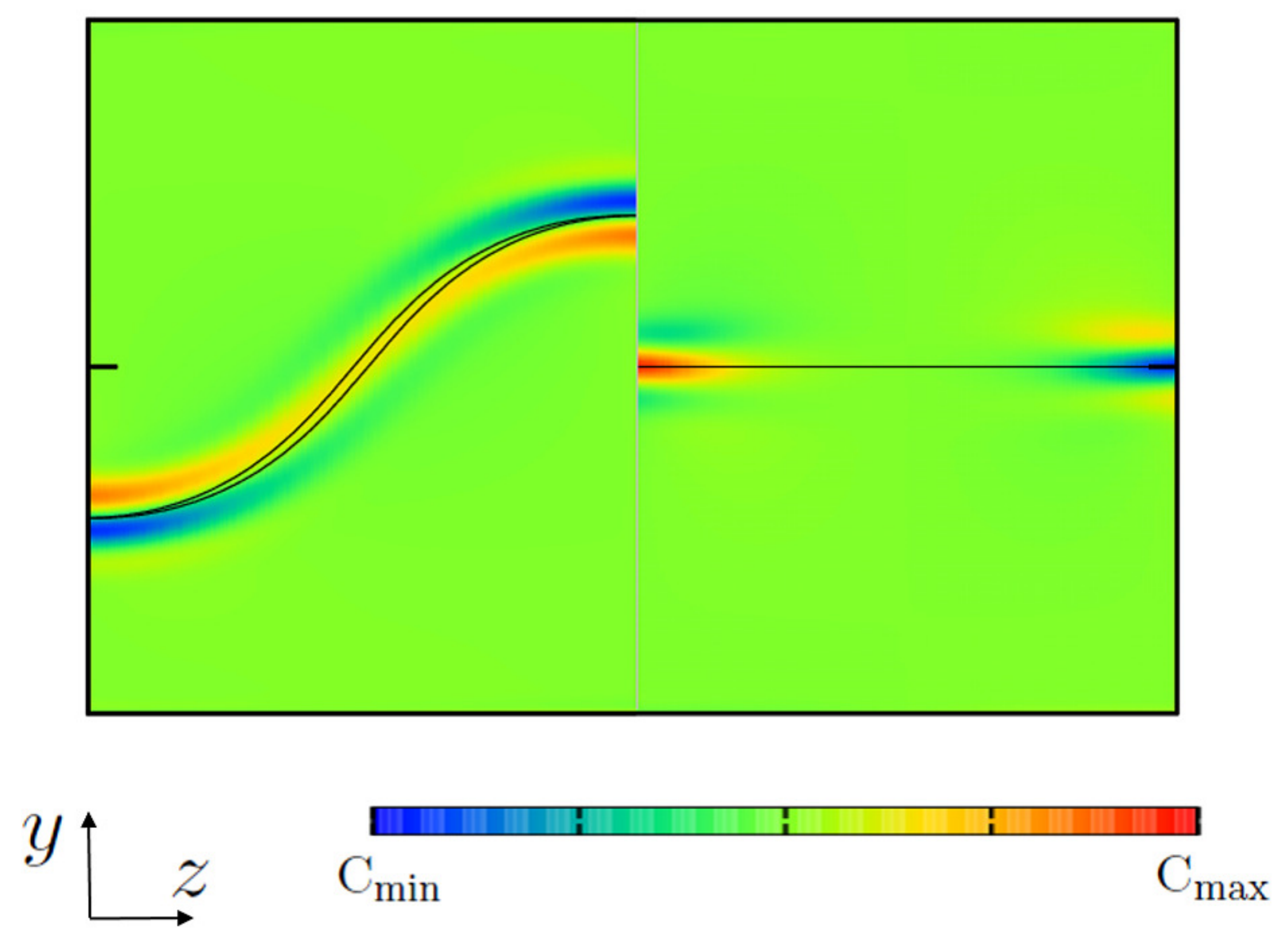} 
\caption{
The streamwise vorticity at $x=0$. The left half is the sinuous mode ($(\text{C}_{\text{min}},\text{C}_{\text{max}})=(-0.05,0.05)$, the same solution as the right panel of figure 2); the right half is the mirror-symmetric mode ($(\text{C}_{\text{min}},\text{C}_{\text{max}})=(-0.08,0.08)$, the same solution as figure 4). The resonant curves are indicated by the black solid curves.
}
\label{fig:longwave}
\end{figure}

\begin{figure}
\centering
\includegraphics[scale=1.1]{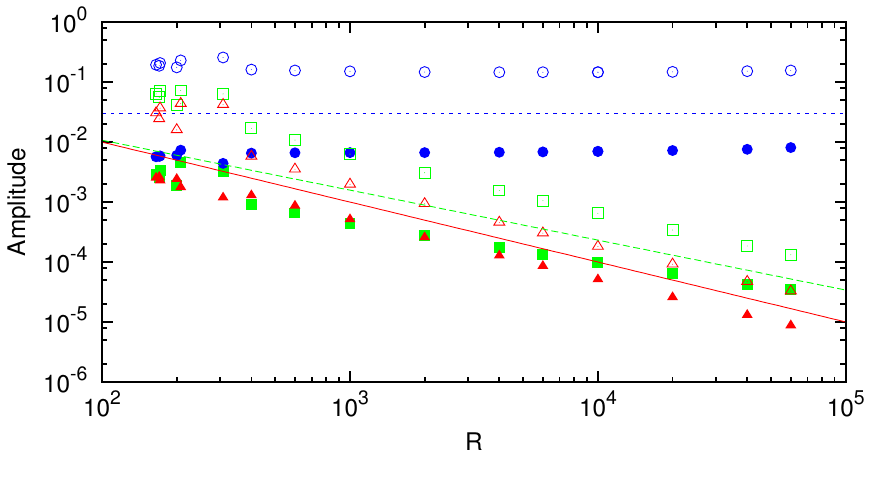} \\
\includegraphics[scale=1.1]{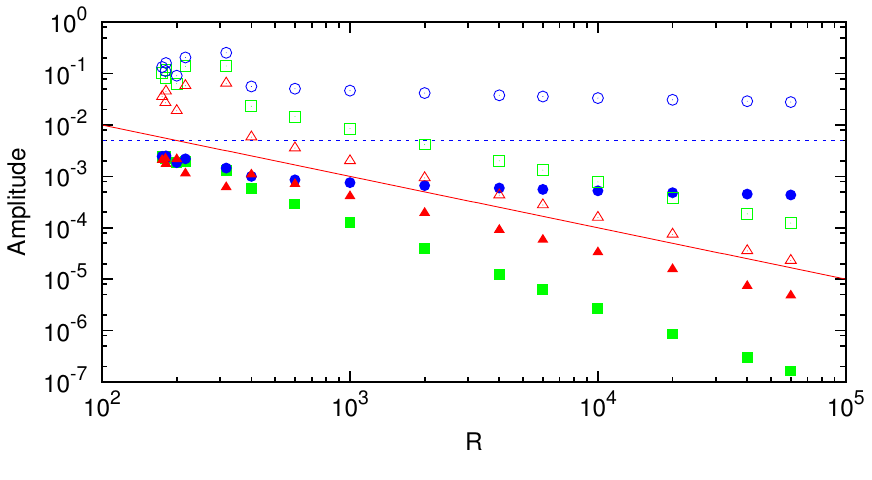} 
\caption{The amplitudes of the solutions subject to the uniform spanwise magnetic field with $B_0=2/R$.
Upper panel: the sinuous mode, lower panel: the mirror-symmetric mode. The open symbols are the hydrodynamic component, the filled symbols are the magnetic component. The red triangles, blue circles and green squares are the roll, streak, and wave amplitudes, respectively. The slope of the blue dotted, red solid, and green dashed lines are $0, -1$, and $-5/6$, respectively.}
\label{fig:longwave}
\end{figure}

For these two types of solutions, we can check the asymptotic scaling varying the Reynolds number.
\textcolor{black}{To} measure the size of the roll-streak-wave component, it is convenient to define the amplitude by the square root of the hydrodynamic streak, magnetic streak, hydrodynamic roll, magnetic roll, hydrodynamic wave and magnetic wave energies defined below:
\begin{eqnarray*}
\frac{1}{2} \langle (\overline{u}-U_b)^2 \rangle, \qquad
\frac{1}{2} \langle (\overline{a}-A_b)^2 \rangle, \qquad
\frac{1}{2} \langle \overline{v}^2+(\overline{w}-W_b)^2 \rangle, \qquad
\frac{1}{2} \langle \overline{b}^2+(\overline{c}-C_b)^2 \rangle,\\
\frac{1}{2} \langle (u-\overline{u})^2+(v-\overline{v})^2+(w-\overline{w})^2 \rangle,\qquad
\frac{1}{2} \langle (a-\overline{a})^2+(b-\overline{b})^2+(c-\overline{c})^2 \rangle,
\end{eqnarray*}
where,
\begin{eqnarray}
\langle~\rangle=\frac{1}{2}\int^{1}_{-1}\overline{\overline{(~)}}dy
\end{eqnarray}
represents the volumetric average over the computational domain.
Note, only the inner wave occurring within the thickness $O(R^{-1/3})$ contributes to the wave amplitude to \textcolor{black}{the} leading order. 
For type 1, the scaling of the amplitude can be estimated as $[(R^{-2/3})^2\times R^{-1/3}]^{1/2}=R^{-5/6}$, whilst the amplitude of type 2 is $[(R^{-5/6})^2\times R^{-1/3}]^{1/2}=R^{-1}$, as summarised in table 2.

Figure 6 shows the dependence of the amplitudes on the Reynolds number. 
The upper and lower panels correspond to the sinuous and mirror-symmetric modes, respectively. 
The spanwise uniform magnetic field with $B_0=2/R$ is used to support the $O(R^{-1})$ roll components shown by the red triangles. 
As predicted by the theory, the streak parts of the solution shown by the blue circles tend \textcolor{black}{towards the} constants for \textcolor{black}{the} large Reynolds numbers, thereby proving the $O(R^0)$ magnetic field generation \textcolor{black}{through a} much smaller external magnetic field. 
The scaling of the wave amplitude, shown by the green squares, is less clear \textcolor{black}{within} this range of $R$.
The wave amplitudes of the mirror-symmetric mode, formally being type 2, drops slightly faster than the sinuous mode of type 1; this is consistent with what has been shown \textcolor{black}{by} Deguchi \textcolor{black}{(2019)}. 
However, the slopes of the numerical results do not perfectly match the theory. 
The slow convergence of the mirror-symmetric mode is not surprising, \textcolor{black}{since} it is known that the asymptotic convergence is not very good even for the purely hydrodynamic case (Deguchi \& Hall 2014). (That slow convergence of the flow seems to typically appear when there \textcolor{black}{exist} two or more possible asymptotic regimes; see Deguchi \& Hall 2015; Ozcakir et al. 2016). 
\textcolor{black}{Alternately,}  the slow convergence of the sinuous mode seems to be due to the different mechanism. 
\textcolor{black}{Recall, that to have the type 1 interaction, the distance between the two resonant curves should be larger than the dissipative layer thickness. In figure 5, this is not satisfied. The reason is that the hydrodynamic and magnetic streak fields have the same symmetry and similar shape; see the right panels in figure 1. 
However, for higher Reynolds numbers, the situation might be different, since the dissipative layer thickness becomes thinner. The distance between the two singular curves may be unchanged as the hydrodynamic and magnetic streaks already become insensitive to the Reynolds number. 
We thus need higher Reynolds number results to conclude this issue, but unfortunately numerical computation becomes difficult due to the demand of high resolution.}

\section{\textcolor{black}{Nonlinear MHD states without any external magnetic field: emergence of dynamo states}}

\subsection{Homotopy using the streamwise magnetic field}
\begin{figure}
\centering
\includegraphics[scale=1.1]{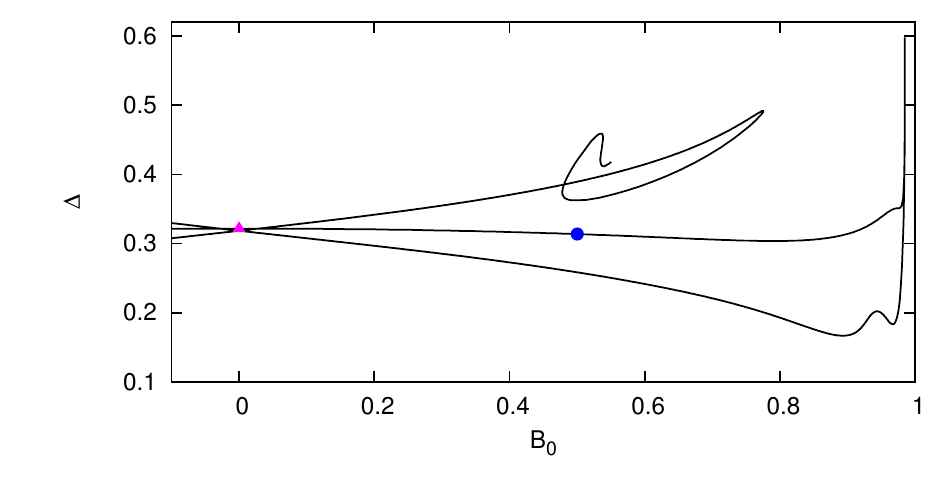} \\
\includegraphics[scale=1.]{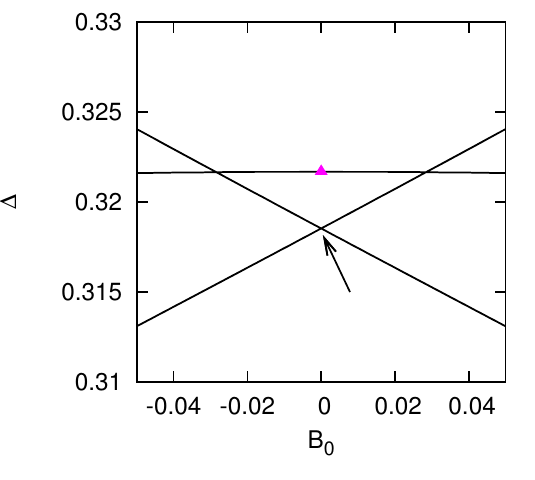} 
\includegraphics[scale=1.]{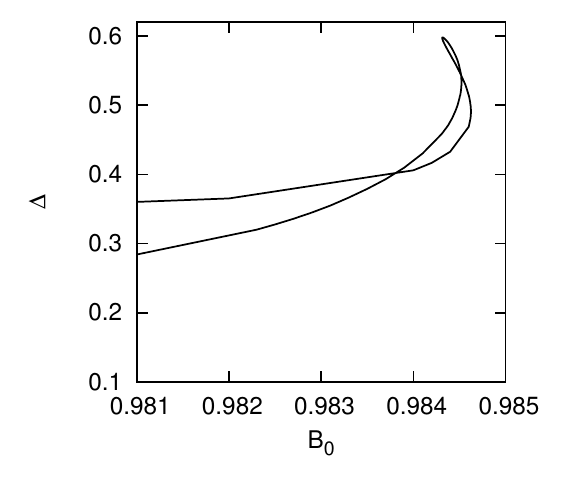} 
\caption{The bifurcation diagram for the external field $A_b=B_0y,C_b=0$. The lower two panels are the magnified plot of the upper panel.
There are three solutions at $B_0=0$: the magenta triangle is the NBC solution, whilst the other two indicated by the arrow are the S$^3$ dynamo solutions. The solution at the filled circle corresponds to the visualisation in the left panels of figures 1,2. }
\label{fig:longwave}
\end{figure}

\begin{figure}
\centering
\includegraphics[scale=0.46]{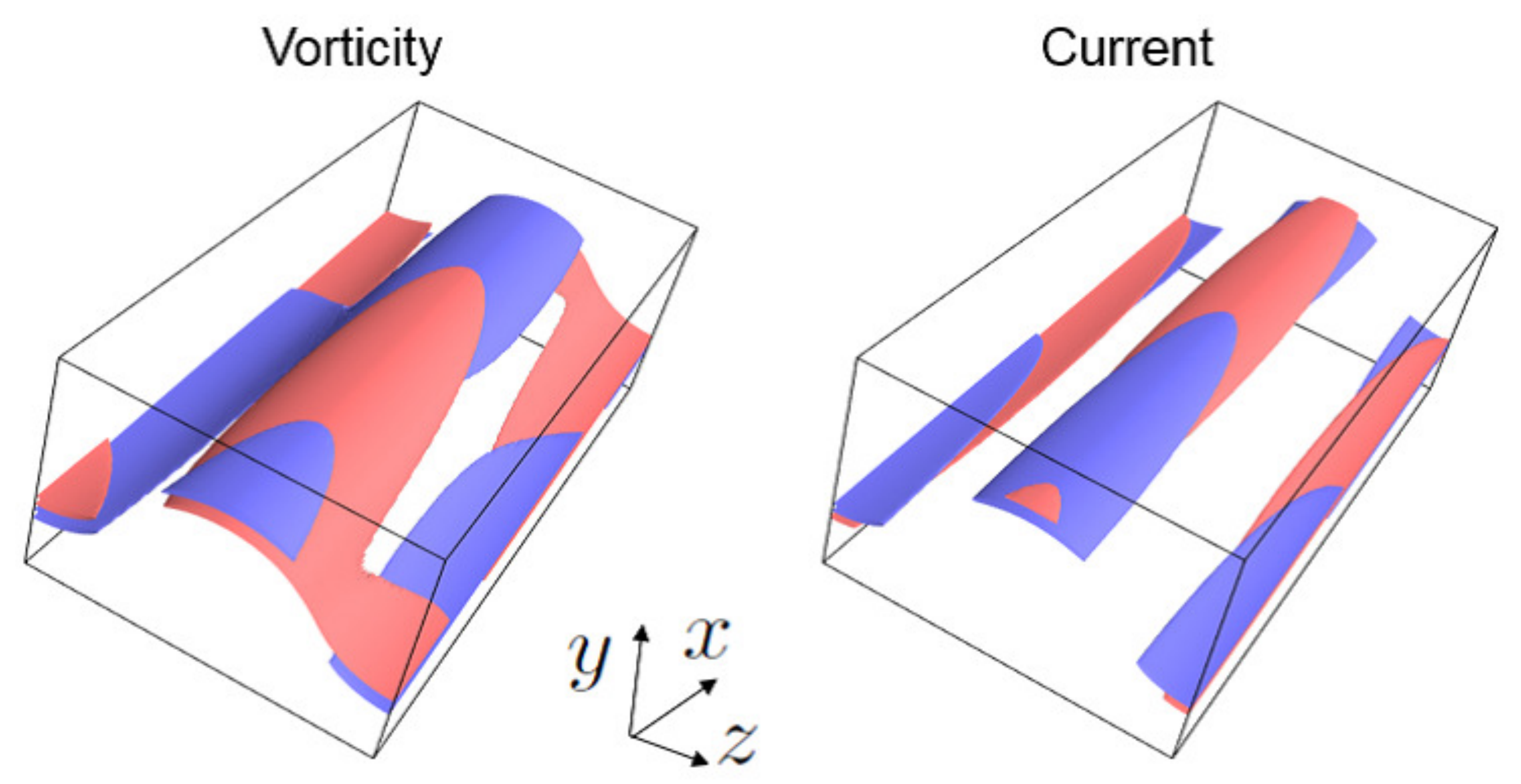} 
\caption{The 50\% streamwise vorticity and current of the S$^3$ dynamo solution in figure 7. Red: positive, blue: negative.}
\label{fig:longwave}
\end{figure}

\begin{figure}
\centering
\includegraphics[scale=0.35]{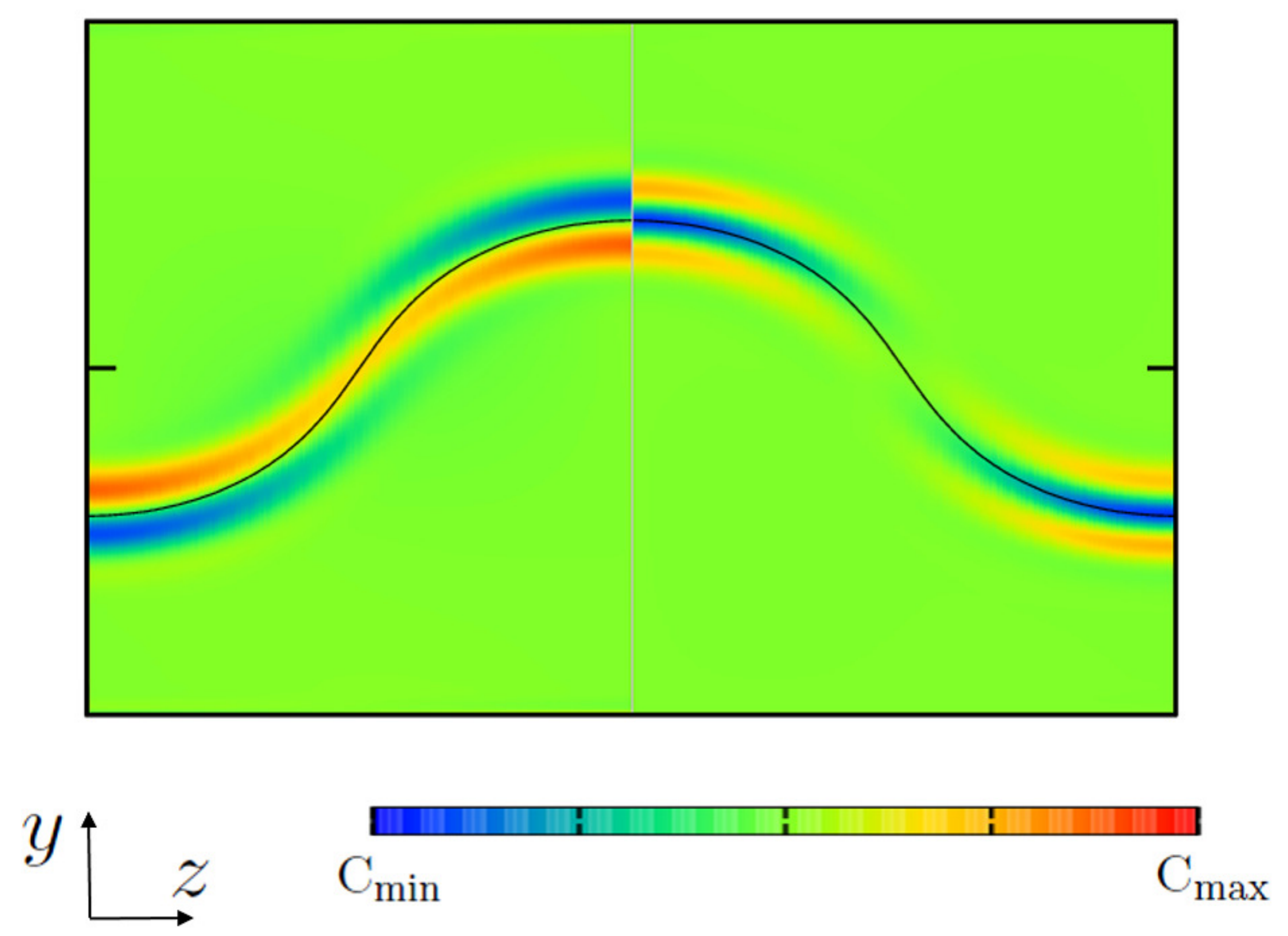} 
\caption{The streamwise vorticity (left half, $(\text{C}_{\text{min}},\text{C}_{\text{max}})=(-0.04,0.04)$) and the current (right half, $(\text{C}_{\text{min}},\text{C}_{\text{max}})=(-0.05,0.05)$) of the S$^3$ dynamo at $x=0$. The same solution as figure 8.
The resonant curve is indicated by the solid black curve.
}
\label{fig:longwave}
\end{figure}

In the previous section, the nonlinear solutions lose all the magnetic \textcolor{black}{fields} as $B_0 \rightarrow 0$, and become purely hydrodynamic solutions. This is consistent with the vortex/Alfv\'en wave interaction theory, where the leading order magnetic field cannot survive at this limit. 
However, further numerical investigations reveal the induced magnetic field does not always vanish at the zero external magnetic field limit. This section is devoted to the analysis of such self-sustained shear-driven dynamos \textcolor{black}{(S$^3$ dynamos).}

\textcolor{black}{To} find the S$^3$ dynamos, we consider \textcolor{black}{a} homotopy continuation of the solution branch using \textcolor{black}{a} different type of external magnetic field. 
Figure 7 shows the continuation of the solution branch when the streamwise magnetic field of \textcolor{black}{the} odd function, $A_b=B_0y, C_b=0$, is used (corresponding to the first case (\ref{sym1})). 
Again, we begin the calculation from the NBC solution indicated by the magenta triangle. 
The solution branch can be continued by gradually increasing $B_0$ until it reaches the turning point, around $B_0=0.9846$. The blue filled circle on the branch corresponds to the visualisation in the left panels of figures 1 \textcolor{black}{and} 2.
\textcolor{black}{After} the turning point, the branch returns to the $B_0=0$ limit as the previous case, but \textcolor{black}{in a} slightly different manner. 

Here, we note that by symmetry, the bifurcation diagram must be symmetric with respect to $B_0=0$. 
\textcolor{black}{Thus, the NBC solution branch continued towards negative $B_0$ must also return to the same point, but the two branches do not coincide unless $B_0=0$, unlike in figure 3.} 
The appearance of the two degenerated solutions at the unforced limit is the signature of the magnetic field production there. 
At $B_0=0$, the symmetry (\ref{symgamma}) ensures that if a S$^3$ dynamo exists, there should be another S$^3$ dynamo of the same velocity but opposite polarity. Therefore, if that branch is continued from S$^3$ dynamos for finite $B_0$, the symmetry of the magnetic field becomes imperfect, and hence, two distinct branches should appear.
On the other hand, such imperfection does not occur when the branch is continued from \textcolor{black}{the} purely hydrodynamic solutions.
After the crossing point, the branch behaves \textcolor{black}{in a} rather complicated way \textcolor{black}{such} the computation is terminated in figure 7.

The three-dimensional structure of one of the S$^3$ dynamos obtained is shown in figure 8. The structure of the solution is similar to the externally-forced solution seen in the left panels in figure 2, \textcolor{black}{where} the symmetry (\ref{sym1}) is preserved. 
The vortex and current sheets in the flow \textcolor{black}{are formed by} the resonant absorption. \textcolor{black}{However,} the mechanism of \textcolor{black}{the fine structure there} is different from the previous cases, as \textcolor{black}{will be explained} shortly.

Varying $R$, the S$^3$ dynamo solution can be traced back to its origin at the saddle-node as the NBC branch, as shown in figure 10.
\textcolor{black}{The} solutions we have seen in figure 7 at $B_0=0$ are the lower branch states.
The saddle-node point of the S$^3$ dynamo solution, shown by the solid curve, sits at \textcolor{black}{a} slightly larger Reynolds number ($R\approx205$) than that of the NBC branch ($R\approx 162$). 
It has been repeatedly shown in the hydrodynamic community that the unstable invariant solutions are the \textcolor{black}{precursors} of turbulence (Gibson et al. (2009), Kreilos \& Eckhardt (2012)). 
Thus the bifurcation diagram shown in figure 10 suggests that to observe the self magnetic field generation, we need \textcolor{black}{a} larger Reynolds number than the critical Reynolds number for the hydrodynamic turbulence. This \textcolor{black}{is} consistent with the recent numerical simulation by Neumann \& Blackman (2017).

\begin{figure}
\centering
\includegraphics[scale=1.1]{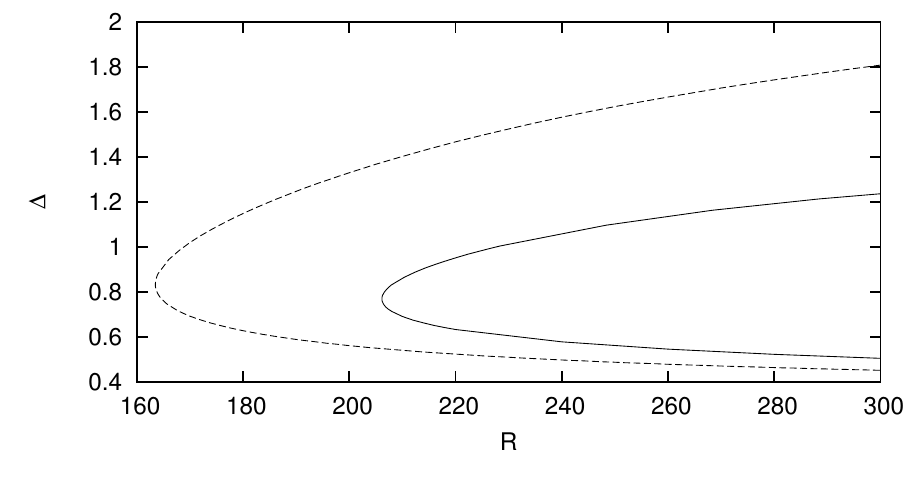} 
\caption{The bifurcation diagram of plane Couette flow with no external magnetic field. 
\textcolor{black}{The S$^3$ dynamo and the NBC solution branches are shown by the solid and dashed curves, respectively.}
}
\label{fig:longwave}
\end{figure}

\subsection{Asymptotic scaling of the self-sustained shear driven dynamos}

Now, let us examine the large Reynolds number asymptotic fate of the S$^3$ dynamos. 
We use the self-consistent matched asymptotic theory proposed by Deguchi \textcolor{black}{(2019)},  \textcolor{black}{partially} motivated by the numerical results in the last section. The ability of the theory to predict the finite Reynolds numbers results will be tested against the numerical invariant solutions.

In view of the results in section 3, one might think the generation of the magnetic field at large Reynolds numbers \textcolor{black}{contradicts} the vortex/Alfv\'en wave interaction theory. 
However, the theory only inhibits the presence of the magnetic field of the order shown in table 2; \textcolor{black}{hence,} smaller magnetic fields could be maintained. 
\textcolor{black}{The difficulty in formulating the asymptotic theory with a smaller magnetic field is that the dynamos tend to become kinematic, i.e. the magnetic field is completely governed by the linear equations.} In \textcolor{black}{such a} case, the magnetic field amplitude is undetermined, so we are unsure if there is \textcolor{black}{a} saturated magnetic field generation. 

The asymptotic theory proposed \textcolor{black}{by} Deguchi \textcolor{black}{(2019)} has overcome this difficulty showing the presence of the nonlinear magnetic effects \textcolor{black}{by the resonant absorption mechanism}, now formally occurring when $\overline{u}-s$ vanishes. 
\textcolor{black}{There is only one resonant layer in the flow because to the leading order, the instability wave is purely hydrodynamic. 
The hydrodynamic wave amplitude behaves like $n^{-1}$ near the resonant layer, where $n$ is the distance to the resonant location, similar to the vortex/Alfv\'en wave interaction theory. 
However, the crucial difference is that the magnetic wave produced at the next order behaves like $n^{-2}$ and is thus much amplified near the resonant curve. }
\textcolor{black}{This} discovery implies that the hydrodynamic and magnetic waves within the dissipative layer of thickness $\delta=R^{-1/3}$ becomes comparable in size if the outer magnetic field size is chosen to be smaller by $O(\delta)$ compared with the hydrodynamic counterpart; see table 3. 
With that choice of scaling, the inner wave Maxwell stress driving the nonlinear magnetic feedback effect to the hydrodynamic flow  has the same magnitude as the inner wave Reynolds stress.
The vortex and current sheets seen in figure 8 prove the existence of \textcolor{black}{the} strong amplification due to the resonance;
see figure 9 where the flow at $x=0$ is shown with the resonant curve.

\begin{table}
  \begin{center}
    \begin{tabular}{cccccc}
    ~~~~~~~~~& Streak &  Roll & Outer wave & Inner wave & Wave amplitude \\
    Hydrodynamic & $O(R^0)$ & $O(R^{-1})$ & $O(R^{-7/6})$ & $O(R^{-5/6})$ & $O(R^{-1})$ \\ 
    Magnetic & $O(R^{-1/3})$ & $O(R^{-4/3})$ & $O(R^{-3/2})$ & $O(R^{-5/6})$ & $O(R^{-1})$ 
    \end{tabular}
  \end{center}
\caption{The scaling of the S$^3$ dynamo states \textcolor{black}{obtained in the asymptotic theory by Deguchi (2019).}} 
\label{sample-table}
\end{table}

\begin{figure}
\centering
\includegraphics[scale=1.1]{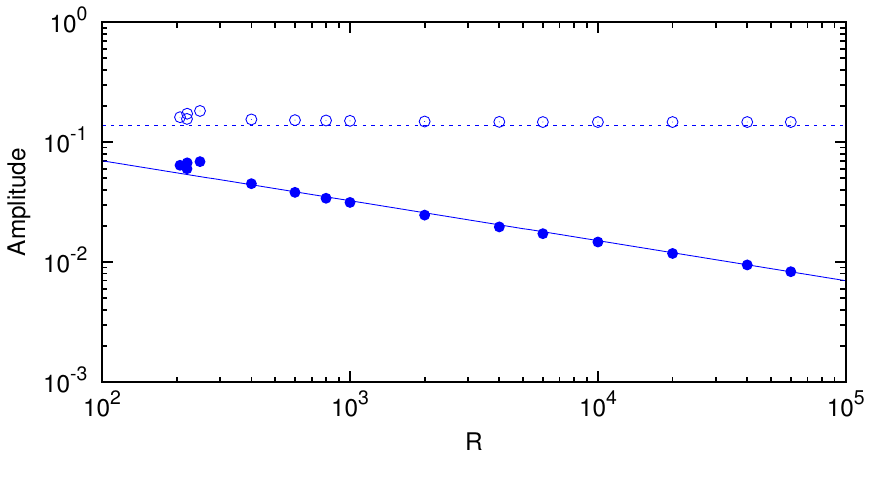} \\
\includegraphics[scale=1.1]{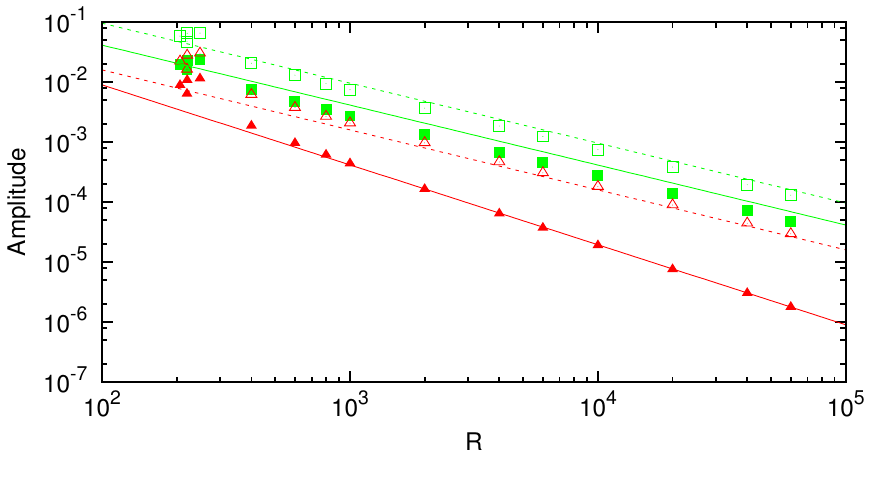} 
\caption{The comparison of the S$^3$ dynamo solutions produced by the MHD equations (\ref{MHD}) and the hybrid equations (\ref{hybrid}).
The points are the MHD results at \textcolor{black}{the} finite Reynolds numbers (the definition is the same as figure 6). 
The lines are the hybrid result based on the asymptotic theory \textcolor{black}{by} Deguchi \textcolor{black}{(2019).} The dashed lines are used for the hydrodynamic parts, and the solid lines correspond to the magnetic parts.
}
\label{fig:longwave}
\end{figure}

\textcolor{black}{As per} Deguchi \textcolor{black}{(2019),} the asymptotic closure was derived by fully scaling out all the Reynolds number dependences of the flow. The solutions of this system enable  
the quantitative comparison of asymptotic results with the finite Reynolds number numerical solutions.
The analytic solution of the asymptotic system is unfortunately not available, because it is still complicated partial differential equations.
\textcolor{black}{The} numerical solutions of the asymptotic system are again not easily found, because we must appropriately treat the singularity at the resonant layer, whose position is a priori unknown.

The main concept of the hybrid approach used in Blackburn et al. (2013) is to avoid the singularity by retaining some dissipative effects in the reduced problem.
More precisely, to derive the hybrid system, we neglect a term when it does not cause any leading order contribution in `all' asymptotic regions (the viscosity and resistivity are retained because these are the leading order effects in the dissipative layer). 
In this sense, the hybrid system is considered to be an intermediately reduced system between the full and the asymptotic system.

In view of the asymptotic theory formulated in Deguchi \textcolor{black}{(2019)}, the hybrid system of the S$^3$ dynamos should possess the following characters:
\begin{enumerate}
\item Only one streamwise Fourier mode contributes to the leading order wave, 
because it should be proportional to the neutral eigensolution of the linear streak stability problem.
\item The effect of the roll in the wave equations is so small that it is negligible everywhere.
\item In the streak equations, the wave components do not cause any leading order effect.
\item In the hydrodynamic part of the equations, all magnetic effects are negligible, except for the wave self-interaction terms (the wave Reynolds and Maxwell stresses) that cause the leading order effect within the dissipative layer.
\end{enumerate}
\textcolor{black}{From (a), again we can express the flow field as (\ref{expincomp}).
Substituting (\ref{expincomp}) to (\ref{MHD}) and then neglecting terms which are negligible everywhere according to the asymptotic theory,} the high Reynolds number travelling waves propagating in the $x$ direction are approximately governed by the vortex equations
\begin{subequations}\label{hybrid}
\begin{eqnarray}
~[(\overline{v}\partial_y +\overline{w}\partial_z)-R_f^{-1}\triangle_2]
\left[ \begin{array}{c} \overline{U}\\ \overline{v}\\ \overline{w} \end{array} \right]
+\left[ \begin{array}{c} 0\\ \overline{q}_y\\ \overline{q}_z \end{array} \right]
=\nonumber \\
-\left[ \begin{array}{c} U_b'\overline{v} \\ 
\{ (|\widetilde{v}|^2-|\widetilde{b}|^2 )_y +( \widetilde{v} \widetilde{w}^* -\widetilde{b} \widetilde{c}^*)_z \}+\text{c.c.} \\ 
\{  (|\widetilde{w}|^2-|\widetilde{c}|^2 )_z + ( \widetilde{v} \widetilde{w}^* -\widetilde{b} \widetilde{c}^*)_y \}+\text{c.c.}
\end{array} \right],~~~\label{invmo}
\\
~[(\overline{v}\partial_y +\overline{w}\partial_z)-R_f^{-1}P_m^{-1}\triangle_2]
\left[ \begin{array}{c} \overline{a}\\ \overline{b}\\ \overline{c} \end{array} \right]
-[\overline{b}\partial_y +\overline{c}\partial_z]
\left[ \begin{array}{c} 0\\ \overline{v}\\ \overline{w} \end{array} \right]=\nonumber \\
\left[ \begin{array}{c} U_b' \overline{b}+(\overline{b}\partial_y +\overline{c}\partial_z)\overline{U}\\ (\widetilde{c}\widetilde{v}^*-\widetilde{b}\widetilde{w}^*)_z+\text{c.c.} \\ -(\widetilde{c}\widetilde{v}^*-\widetilde{b}\widetilde{w}^*)_y+\text{c.c.} \end{array} \right],~~~\label{invind}
\\
\overline{v}_y+\overline{w}_z=0, \qquad \overline{b}_y+\overline{c}_z=0, ~~~~~\label{invcontv}
\end{eqnarray}
and the wave equations 
\begin{eqnarray}
\left \{
(\overline{u}-s)i\alpha 
\left[ \begin{array}{c} \widetilde{u}\\ \widetilde{v}\\ \widetilde{w} \end{array} \right] 
+
\left[ \begin{array}{c} \widetilde{v}\overline{u}_y+\widetilde{w}\overline{u}_z\\ 0\\ 0 \end{array} \right]
\right \} 
+\left[ \begin{array}{c} i\alpha \widetilde{q} \\ \widetilde{q}_y \\ \widetilde{q}_z \end{array} \right]=R_f^{-1}\triangle 
\left[ \begin{array}{c} \widetilde{u}\\ \widetilde{v}\\ \widetilde{w} \end{array} \right],~~~~~\label{Cwavemo}
\\
\left \{
(\overline{u}-s)i\alpha
\left[ \begin{array}{c} \widetilde{a}\\ \widetilde{b}\\ \widetilde{c} \end{array} \right] 
+
\left[ \begin{array}{c} \widetilde{v}\overline{a}_y+\widetilde{w}\overline{a}_z\\ 0\\ 0 \end{array} \right]
\right \} \hspace{70mm} \nonumber \\
-
\left \{
\overline{a}i\alpha
\left[ \begin{array}{c} \widetilde{u}\\ \widetilde{v}\\ \widetilde{w} \end{array} \right] 
+
\left[ \begin{array}{c} \widetilde{b}\overline{u}_y+\widetilde{c}\overline{u}_z\\ 0\\ 0 \end{array} \right]
\right \}
=R_f^{-1}P_m^{-1}\triangle 
\left[ \begin{array}{c} \widetilde{a}\\ \widetilde{b}\\ \widetilde{c} \end{array} \right],~~~\label{Cwaveind}\\
i\alpha \widetilde{u}+\widetilde{v}_y+\widetilde{w}_z=0,\qquad i\alpha \widetilde{a}+\widetilde{b}_y+\widetilde{c}_z=0,~~~~~\label{Cwavecont}
\end{eqnarray}
\end{subequations}
where,  $R_f$ is the fictitious Reynolds number. 
The numerical solution of this hybrid system can be found by using similar numerical method \textcolor{black}{in} section 2. 
For \textcolor{black}{a} large enough $R_f$, the scaled solution becomes independent of \textcolor{black}{the} $R_f$ and should match the full asymptotic result as $R_f^{-1}$ represents the strength of the regularisation; here, we \textcolor{black}{have choosen} $R_f=500000$. \textcolor{black}{To} capture the sharp dissipative layer structure, the higher resolution of $(L,M)=(220,40)$ is used in the $y-z$ plane.

Figure 11 compares the amplitude of the finite Reynolds number computations and the asymptotic result. The lines in the figure are the asymptotic results, having the slope \textcolor{black}{and intercept found in table 3 and the hybrid results, respectively.}
The asymptotic prediction for the full numerical solutions shown by the points is remarkably good, even for Reynolds numbers as low as $10^3$.

\section{Conclusion}

The \textcolor{black}{predominantly} shear-driven \textcolor{black}{nonlinear invariant MHD solutions}, which persist for \textcolor{black}{the} asymptotically large Reynolds numbers are found for the first time. 
They are characterised by the roll-streak structure as the vortex/wave interaction or self-sustaining process, but now the similar structure also appears for the magnetic field. 
The streamwise vortex and current tubes corresponding to the roll components are supported by the hydrodynamic and magnetic instability waves. 
The \textcolor{black}{nonlinear MHD solutions} found are subcritical, in the sense that they do not rely on any linear instability of the base flow. 

Consistent to the vortex/Alfv\'en wave interaction theory proposed \textcolor{black}{by} Deguchi \textcolor{black}{(2019)}, the \textcolor{black}{MHD solutions} can be obtained by imposing a spanwise uniform external magnetic field to the known hydrodynamic invariant solutions. 
As shown in section 3, the spanwise magnetic field of $O(R^{-1})$ is chosen \textcolor{black}{such} that it will pour energy into the magnetic roll, which in turn \textcolor{black}{will generate a} much bigger $O(R^0)$ magnetic field through the omega effect. 
We first \textcolor{black}{computed} the sinuous mode \textcolor{black}{solution} branch by continuing the NBC solution, and then found that it connects to another \textcolor{black}{solution} branch associated with the mirror-symmetric mode.
The structures of the numerical solutions are found to be consistent with the theoretical prediction that the generated streamwise magnetic field triggers the Alfv\'en waves, which resonate at the singular curves to produce the strong vortex and current sheets \textcolor{black}{therein}. 
In the computation in section 3, all magnetic \textcolor{black}{fields disappear} in the limit of \textcolor{black}{the} zero external magnetic field.
\textcolor{black}{Therefore, these are not dynamo in usual sense.}


To \textcolor{black}{generate} the self-sustained shear-driven dynamos (S$^3$ dynamos), section 4 \textcolor{black}{has} used the homotopy continuation of the solution branch via the augmented system forced by a spanwise uniform current.
The flow structure of the S$^3$ dynamo solutions is similar to the vortex/Alfv\'en wave interaction states, including the singular wave structure \textcolor{black}{owed} to the absorption. \textcolor{black}{However,} the scaling of it with respect to the Reynolds number is different. 
\textcolor{black}{The generated magnetic field is smaller and of $O(R^{-1/3})$;} hence, it does not contradict the caveat for the vortex/Alfv\'en wave interaction theory. 
The self-consistent asymptotic theory for the S$^3$ dynamos shown by Deguchi \textcolor{black}{(2019)} \textcolor{black}{has led to} the reduced equations that can be solved numerically.
The agreement of the asymptotic solution with the full numerical solutions at finite Reynolds numbers is excellent, even for moderate Reynolds numbers.

\textcolor{black}{
The S$^3$ dynamo solutions are slow dynamos in classical dynamo parlance, i.e. the generated magnetic field becomes asymptotically small as magnetic diffusivity tends to zero. 
Thus, one may think they are likely not highly relevant as an efficient magnetic field generation mechanism in the natural dynamo context. 
However, previous hydrodynamic studies suggest that the S$^3$ dynamo solutions may trigger a more energetic turbulent dynamo action. 
The lower branch states in subcritical shear flows, such as the plane Couette flow, are known to have only a few unstable eigenvalues. 
One of these, called the `edge mode' of instability, governs whether the nearby trajectory moves towards the turbulent attractor (Itano \& Toh 2001; Skufca et al. 2006; Wang et al. 2007; Deguchi \& Hall 2016). 
The unstable manifold associated with that mode interacts with period doubling cascade of energetic upper branch to form turbulent attractor (Gibson et al. 2008; van Veen \& Kawahara 2011; Kreilos \& Eckhardt 2012; Lustro et al. 2019).
The S$^3$ dynamo solution is perhaps weakly unstable and has an edge mode, since it is a lower branch; in fact, it resembles the lower branch NBC solution studied in Wang et al. (2007). 
Moreover, its upper branch state has much stronger magnetic field generation, whose behaviour is apparently unexplained by the asymptotic theory; see figure 12. 
Therefore, the analogy to the hydrodynamic case suggests that the smallness of the magnetic field in the S$^3$ dynamos may actually imply that the asymptotically small magnetic field can stimulate energetic subcritical turbulent dynamo. 
The instability of the lower branch S$^3$ dynamos control the transition in this scenario, and the asymptotic development of it could be analysed by a similar manner as in Deguchi \& Hall (2016). 
The instability may possibly have some relevance to the plasmoid instabilities typically occurring in the thin current sheet structure (Loureiro et al. 2007). Thus, it is of interest to look at the scaling of the magnetic perturbations with $R_m$ to see if reconnection in the large Lundquist number regime of fast MHD reconnection is possible for these solutions. 
}

\begin{figure}
\centering
\includegraphics[scale=1.1]{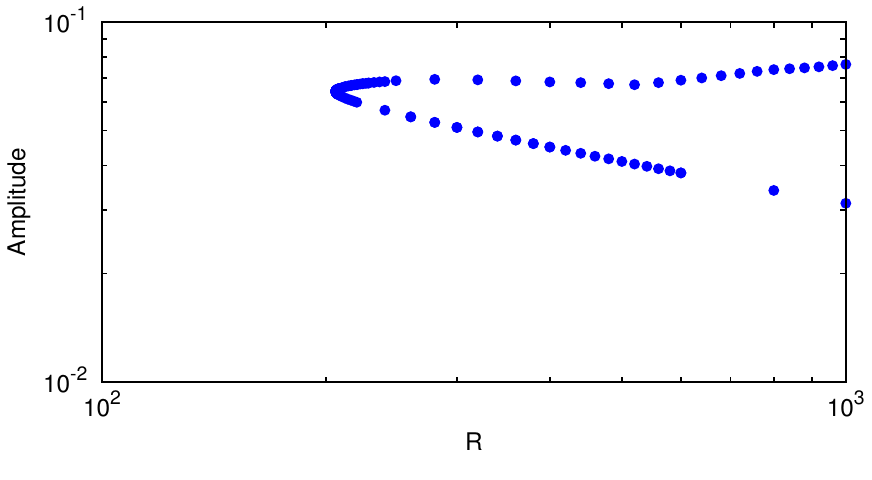} 
\caption{The magnetic streak amplitude of the dynamo solution. 
The lower branch converges to the asymptotic theory by Deguchi (2019) as seen in figure 11. The upper branch state generates much larger magnetic field.
}
\label{fig:longwave}
\end{figure}

The \textcolor{black}{overall} interaction mechanism of our asymptotic theory is independent of specific flow configurations; \textcolor{black}{thus,} our theory would serve a new perspective to various areas of research.
For example, the interaction of Alfv\'en wave and solar atmosphere has been studied by various model approaches (Antolin \& Shibata 2010; van Ballegooijen 2011).
The asymptotic states concerned here and \textcolor{black}{in} Deguchi \textcolor{black}{(2019)} could be viewed as an extension of the resonant absorption theories developed in solar physics (Sakurai et al. 1991; Goossens et al. 1992).
The direct observational evidence of the resonant absorption in the Sun's atmosphere has recently been reported (Okamoto et al. 2015) revealing that the thin hot current sheet produced by the absorption in coronal loops and solar flare would be key to unlocking the long standing coronal heating problem. 
Interestingly, the presence of the thread like structures reminiscent of the roll-streak structures in the solar atmosphere \textcolor{black}{has been} reported. 
The theory presented in this paper may have some connection to the interaction between the inhomogeneous prominence thread and \textcolor{black}{the} Alfv\'en waves via resonant absorption (Arregui et al. 2008; Soler et al. 2012). 

In this paper, incompressibility is assumed. \textcolor{black}{However, in general,} the solar atmosphere is highly compressible. \textcolor{black}{The} previous resonant absorption theories by Sakurai et al. (1991) and Goossens et al. (1992) studied both \textcolor{black}{the} Alfv\'en and cusp wave resonant points, where the latter only occurs for the compressible cases. 
The effect of compressibility can be fairly easily introduced in the nonlinear three-dimensional framework presented here, using the knowledge of the previous resonant absorption theories. 
The extended theory for the compressible case, which should tell the singular variation of temperature around the dissipative layer, will be reported shortly in the future. 

\textcolor{black}{
The study of the $P_m$ dependence of the $S^3$ dynamo solutions is another interesting follow-up in the context of understanding the relevance of them to astrophysical and planetary dynamos, where $P_m$ is usually very low or very large depending on the object. 
The asymptotic theory formulated in Deguchi (2019) works when $P_m$ is $O(R^0)$ or larger and can be thus used to express extremely large $P_m$ flows. 
Conversely, the theory breaks down when $P_m$ gets asymptotically small, because the diffusive terms dominates the induction equations; hence, a nonlinear state seems unrealisable unless a very strong external magnetic field is applied. 
However, as long as $P_m$ is a constant, no matter how small this is, there is no problem in using the asymptotic theory. 
Further parameter search will reveal a critical value of $P_m$ below which numerical solutions of the asymptotic problem do not exist. }


\textcolor{black}{This work is supported by Australian Research Council Discovery Early Career Re- searcher Award DE170100171. The helpful comments of the referees should be gratefully acknowledged.}

\appendix

\section{The nonlinear terms in the fluctuation parts of the equations}
The nonlinear terms in \textcolor{black}{(\ref{flucequations}) and (\ref{meanequations})} can explicitly be computed as follows.
\textcolor{black}{
\begin{subequations}
\begin{eqnarray*}
\mathcal{N}^{(1)}(m,n,m_1,n_1)&=&-L_1L(L_+-L_2)(\widehat{\phi}_1\widehat{\phi}_2'-\widehat{f}_1\widehat{f}_2')\nonumber \\
&&+[L_+(2L_+ +L_2)-L_1L_2](\widehat{\phi}_1'\widehat{\phi}_2''-\widehat{f}_1'\widehat{f}_2'')\nonumber \\
&&-L_1(L_+ +L_2)(\widehat{\phi}_1\widehat{\phi}_2'''-\widehat{f}_1\widehat{f}_2''')\nonumber \\
&&+L_-(2L_+ +L_1)(\widehat{\phi}_1''\widehat{\psi}_2-\widehat{f}_1''\widehat{g}_2)-L_-L_1(\widehat{\phi}_1\widehat{\psi}_2''-\widehat{f}_1\widehat{g}_2'')\nonumber \\
&&+2L_-L_+(\widehat{\phi}_1'\widehat{\psi}_2'-\widehat{f}_1'\widehat{g}_2')-L_-L_1 L(\widehat{\phi}_1\widehat{\psi}_2-\widehat{f}_1\widehat{g}_2)\nonumber \\
&&-2L_-^2(\widehat{\psi}_1\widehat{\psi}_2'-\widehat{g}_1\widehat{g}_2'),\\
\mathcal{N}^{(2)}(m,n,m_1,n_1)&=&L_1L_- (\widehat{\phi}_1 \widehat{\phi}_2''-\widehat{f}_1\widehat{f}_2'')
+(L_+ +L_1)L_2(\widehat{\phi}_1'\widehat{\psi}_2-\widehat{f}_1'\widehat{g}_2)\nonumber \\
&&-L_1(L_+ +L_2)(\widehat{\phi}_1\widehat{\psi}_2'-\widehat{f}_1\widehat{g}_2')
-L_- L_2(\widehat{\psi}_1\widehat{\psi}_2-\widehat{g}_1\widehat{g}_2),\\
\mathcal{N}^{(3)}(m,n,m_1,n_1)&=&
-(L_+ +L_1)L_2(\widehat{\phi}_1'\widehat{f}_2-\widehat{f}_1'\widehat{\phi}_2)
+L_- L_2(\widehat{g}_1\widehat{\phi}_2-\widehat{\psi}_1\widehat{f}_2),\\
\mathcal{N}^{(4)}(m,n,m_1,n_1)&=&
L_1L_- (\widehat{\phi}_1 \widehat{f}_2''-\widehat{f}_1\widehat{\phi}_2'')-2L_+ L_-\widehat{\phi}_1'\widehat{f}_2'\nonumber \\
&&+\{(L_+ +L_1)L_2-2C^2\}(\widehat{\phi}_1'\widehat{g}_2-\widehat{f}_1'\widehat{\psi}_2)\nonumber \\
&&-L_1(L_+ +L_2)(\widehat{\phi}_1\widehat{g}_2'-\widehat{f}_1\widehat{\psi}_2')-L_-(L_+ +L_2)(\widehat{\psi}_1\widehat{g}_2-\widehat{g}_1\widehat{\psi}_2),~~~~\\
\mathcal{N}_0^{(1)}(m_1,n_1)&=&L_1[m_2\alpha(\widehat{f}_1\widehat{f}_2'-\widehat{\phi}_1\widehat{\phi}_2')'
+n_2\beta(\widehat{f}_1\widehat{g}_2-\widehat{\phi}_1 \widehat{\psi}_2)'],\\
\mathcal{N}_0^{(2)}(m_1,n_1)&=&L_1[n_2\beta(\widehat{f}_1\widehat{f}_2'-\widehat{\phi}_1\widehat{\phi}_2')'-m_2\alpha(\widehat{f}_1\widehat{g}_2-\widehat{\phi}_1\widehat{\psi}_2)'],\\
\mathcal{N}_0^{(3)}(m_1,n_1)&=&L_1[m_2\alpha(\widehat{f}_1\widehat{\phi}_2'-\widehat{\phi}_1 \widehat{f}_2')'+n_2\beta(\widehat{f}_1\widehat{\psi}_2-\widehat{\phi}_1 \widehat{g}_2)'], \\
\mathcal{N}_0^{(4)}(m_1,n_1)&=&L_1[n_2\beta(\widehat{f}_1\widehat{\phi}_2'-\widehat{\phi}_1 \widehat{f}_2')'
-m_2\alpha(\widehat{f}_1\widehat{\psi}_2-\widehat{\phi}_1 \widehat{g}_2)'],
\end{eqnarray*}
\end{subequations}
where
\begin{subequations}
\begin{eqnarray*}
L_1=(m_1\alpha)^2+(n_1\beta)^2,\qquad L_2=(m_2\alpha)^2+(n_2\beta)^2,\\
L_+=(m_1\alpha)(m_2\alpha)+(n_1\beta)(n_2\beta),\qquad L_-=(m_1\alpha)(n_2\beta)-(m_2\alpha)(n_1\beta).
\end{eqnarray*}
\end{subequations}
}

\section{Computation with the streamwise external magnetic field of odd function}
We can find the S$^3$ dynamo solutions by applying the streamwise magnetic field of \textcolor{black}{the} even function to the NBC solution. Figure 13 is the computational result produced by the external field $A_b=B_0,C_b=0$. 
The visualisations at the filled circle shown in middle panels of \textcolor{black}{figures 1 and 2} indicate the solutions have the symmetry (\ref{sym2}).
The solution branch behaves \textcolor{black}{in a} complicated way after the first turning point around $B_0\approx 0.065$; \textcolor{black}{however,} it eventually crosses the $B_0=0$ axis twice. 
At the first crossing point, there are two degenerated solutions being opposite polarity each other. 
Thus they are the S$^3$ dynamo solutions whose symmetry becomes imperfect for $B_0\neq 0$. 
On the other hand, at the second crossing point \textcolor{black}{this} imperfection does not occur; \textcolor{black}{thus,} this is a purely hydrodynamic solution.

\begin{figure}
\centering
\includegraphics[scale=1.1]{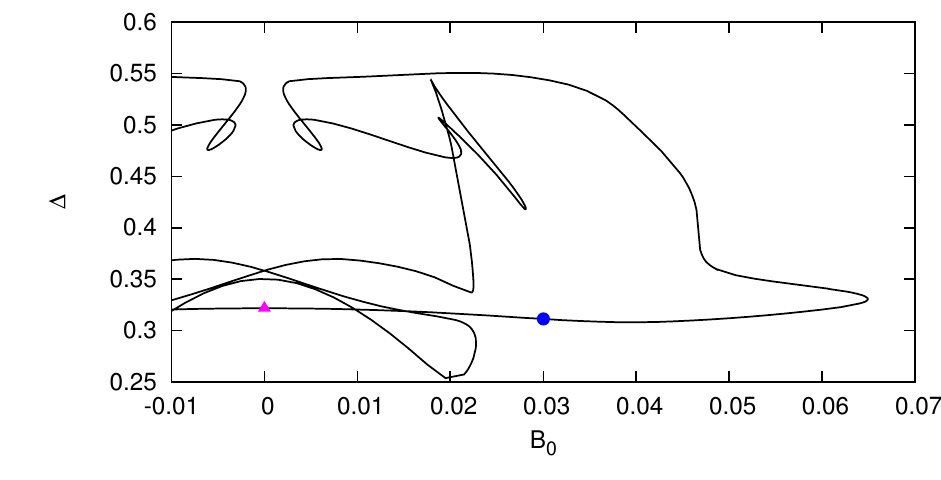} 
\caption{The bifurcation diagram for the external field $A_b=B_0,C_b=0$. The magenta triangle is the NBC solution. The visualisations in the middle panels of \textcolor{black}{figures 1 and 2} correspond to the filled circle.}
\label{fig:longwave}
\end{figure}


\begin{thebibliography}{99}
\bibitem[]{BH} \textsc{Antolin, P. \& Shibata, K.} 2010 The role of torsional Alfv\'en waves in coronal heating. \emph{Astrophys. J.} \textbf{712}, 494--510.

\bibitem[]{BH} \textsc{Arregui, I., Terradas, J. Oliver, R. \& Ballester, J. L.} 2008 Damping of fast magnetohydrodynamic oscillations in quiescent filament threads. \emph{Astrophys. J.} \textbf{682}, L141--L144.

\bibitem[]{BH} \textsc{Balbus, S. A. \& Hawley, J. F.} 1991 A powerful local shear instability in weakly magnetized disks. I. Linear analysis. \emph{Astrophys. J.} \textbf{376}, 214--222.

\bibitem[]{BH} \textsc{Blackburn, H. M., Hall, P. \& Sherwin, S.} 2013 Lower
branch equilibria in Couette flow: the emergence of canonical states for
arbitrary shear flows. \emph{J. Fluid Mech.} \textbf{726}, R2.

\bibitem[]{BH} \textsc{Clever, R. M. \& Busse, F. H.} 1992 Three-dimensional convection in a horizontal fluid layer subjected to a constant shear. \emph{J. Fluid Mech.} \textbf{234}, 511--527.

\bibitem[]{C} \textsc{Cowling, T. G.} 1934 The magnetic fields of sunspots. \emph{Mon. Not. Roy. Astron. Soc.} \textbf{94}, 39--48.

\bibitem[]{D15} \textsc{Deguchi, K.} 2015 Self-sustained states at Kolmogorov microscale. \emph{J. Fluid Mech.} \textbf{781}, R6.

\bibitem[]{D17} \textsc{Deguchi, K.} 2017 Scaling of small vortices in stably stratified shear flows. \emph{J. Fluid Mech.} \textbf{821}, 582--594.
\textcolor{black}{
\bibitem[]{D17} \textsc{Deguchi, K.} 2019 High-speed shear driven dynamos. Part 1. Asymptotic analysis. Accepted to \emph{J. Fluid Mech.} (arXiv:1809.03853)}

\bibitem[]{DH14} \textsc{Deguchi, K. \& Hall, P.} 2014a Canonical exact coherent structures embedded in high Reynolds number flows. \emph{Phil. Trans. R. Soc. Lond.} A \textbf{372}, 20130352, 1--19.

\bibitem[]{DH14} \textsc{Deguchi, K. \& Hall, P.} 2014b The high Reynolds number asymptotic development of nonlinear equilibrium states in plane Couette flow. \emph{J. Fluid Mech.} \textbf{750}, 99--112.

\bibitem[]{DH14} \textsc{Deguchi, K. \& Hall, P.} 2014c Free-stream coherent structures in parallel boundary-layer flows. \emph{J. Fluid Mech.} \textbf{752}, 602--625.

\bibitem[]{DH14} \textsc{Deguchi, K. \& Hall, P.} 2015 Asymptotic descriptions of oblique coherent structures in shear flows. \emph{J. Fluid Mech.} \textbf{782}, 356--367.

\bibitem[]{DH16} \textsc{Deguchi, K. \& Hall, P.} 2016 On the instability of vortex-wave interaction states. \emph{J. Fluid Mech.} \textbf{802}, 634--666.

\bibitem[]{DH16} \textsc{Deguchi, K. \& Walton, A. G.} 2013a Axisymmetric travelling waves in annular sliding Couette flow at finite and asymptotically large Reynolds number. \emph{J. Fluid Mech.} \textbf{720}, 582--617.

\bibitem[]{DH16} \textsc{Deguchi, K. \& Walton, A. G.} 2013b A swirling spiral wave solution in pipe flow. \emph{J. Fluid Mech.} \textbf{737}, R2.

\bibitem[]{DH16} \textsc{Deguchi, K. \& Walton, A. G.} 2018 Bifurcation of nonlinear Tollmien-Schlichting waves in a high-speed channel flow. \emph{J. Fluid Mech.} \textbf{843}, 53--97.

\bibitem[]{DHW} \textsc{Deguchi, K., Hall, P. \& Walton, A. G.} 2013 The emergence of localized vortex--wave interaction states in plane Couette flow. \emph{J. Fluid Mech.} \textbf{721}, 58--85.

\bibitem[]{DDHW} \textsc{Dempsey, L. J., Deguchi, K., Hall, P. \& Walton, A. G.} 2016 Localized vortex/Tollmien-Schlichting wave interaction states in plane Poiseuille flow. \emph{J. Fluid Mech.} \textbf{791}, 97--121.

\bibitem[]{DHW} \textsc{Gibson, J. F., Halcrow, J. \& Cvitanovic, P.} 2008 Visualizing the geometry of state space in plane Couette flow. \emph{J. Fluid Mech.} \textbf{611}, 107--130.

\bibitem[]{DHW} \textsc{Gibson, J. F., Halcrow, J. \& Cvitanovic, P.} 2009 Equilibrium and travelling-wave solutions of plane Couette flow. \emph{J. Fluid Mech.} \textbf{638}, 1--24.
\textcolor{black}{
\bibitem[]{DHW} \textsc{Guseva, A., Hollerbach, R., Willis, A. P. \& Avila, M.} 2017 Dynamo action in a quasi-Keplerian Taylor-Couette flow. \emph{Phys. Rev. Lett.} \textbf{119}, 164501.}

\bibitem[]{GHS92} \textsc{Goossens, M., Hollweg, J. V. \& Sakurai, T.} 1992 Resonant behaviour of MHD waves on magnetic flux tubes. III. Effect of equilibrium flow. \emph{Solar Physics} \textbf{138}, 233--255.

\bibitem[]{DHW} \textsc{Itano, T. \& Generalis, S.} 2009 Hairpin vortex solution in planar Couette flow: a tapestry of knotted vortices. \emph{Phys. Rev. Lett} \textbf{102}, 114501.

\bibitem[]{DHW} \textsc{Itano, T. \& Toh, S.} 2001 The dynamics of bursting process in wall turbulence. \emph{J. Phys. Soc. Japan} \textbf{70}, 703--716.

\bibitem[]{K2012} \textsc{Kawahara, G. \& Kida, S.} 2001 Periodic motion embedded in plane Couette turbulence: regeneration cycle and burst. \emph{J. Fluid Mech.} \textbf{449}, 291--300.

\bibitem[]{K2012} \textsc{Kawahara, G., Uhlmann, M. \& van Veen, L.} 2012 The significance of simple invariant solutions in turbulent flows. \emph{Ann. Rev. Fluid Mech.} \textbf{44}, 203--225.

\bibitem[]{K2012} \textsc{Kreilos, T. \& Eckhardt, B.} 2012 Periodic orbits near onset of chaos in plane Couette flow. \emph{Chaos} \textbf{22}, 047505.

\bibitem[]{K2012} \textsc{Lorenz, E. N. } 1963 Deterministic nonperiodic flow. \emph{J. Atmos. Sci.} \textbf{20}, 130--141.
\textcolor{black}{
\bibitem[]{K2012} \textsc{Loureiro, N. F., Schekochihin, A. A. \& Cowley, S. C. } 2007 Instability of current sheets and formation of plasmodia chains. \emph{Phys. Plasmas} \textbf{14}, 100703.
}
\textcolor{black}{
\bibitem[]{K2012} \textsc{Lustro, J. R. T., Kawahara, G., van Veen, L., Shimizt, M. \& Kokubu, H.} 2019 The onset of transient turbulence in minimal plane Couette flow. \emph{J. Fluid Mech.} \textbf{862}, R2.
}
\textcolor{black}{
\bibitem[]{lin45} \textsc{Marcotte, F. \& Gissinger, C.} 2016 Dynamo generated by the centrifugal instability. \textit{Phys. Rev. Fluids} \textbf{1}, 063602.}

\bibitem[]{K2012} \textsc{Nagata, M. } 1990 Three-dimensional finite-amplitude solutions in plane Couette flow: bifurcation from infinity. \emph{J. Fluid Mech.} \textbf{217}, 519--527.

\bibitem[]{lin45} \textsc{Nauman, F. \& Blackman, E. G.} 2017 Sustained turbulence and magnetic energy in nonrotating shear flows. \textit{Phys. Rev. E} \textbf{95}, 033202.
\textcolor{black}{
\bibitem[]{lin45} \textsc{Nore, C., Guermond, J.-L., Laguerre, R., L\'eorat, J. \& Luddens, F.} 2012 Nonlinear dynamo in a short Taylor-Couette setup. \textit{Phys. Fluids} \textbf{24}, 094106.}

\bibitem[]{lin45} \textsc{Okamoto, T. J., Antolin, P., Pontieu, B. E., Uitenbroek, H., van Doorsselaere, T. \& Yokoyama, T.} 2015 Resonant absorption of transverse oscillations and associated heating in a solar prominence. I. Observational aspects. \textit{Astrophys. J.} \textbf{809}:71, 1--12.

\bibitem[]{lin45} \textsc{Ozcakir, O., Tanveer, S., Hall, P., \& Overman II, E. A. } 2016 Travelling wave states in pipe flow. \textit{J. Fluid Mech.} \textbf{791}, 284--328.

\item \textsc{Hall, P. \& Horseman, N.} 1991 The linear inviscid secondary instability of longitudinal vortex structures in boundary layers. \emph{J. Fluid Mech.} \textbf{232}, 357--375.

\bibitem[]{HS10} \textsc{Hall, P. \& Sherwin, S.} 2010 Streamwise vortices in shear flows: harbingers of transition and the skeleton of coherent structures. \emph{J. Fluid Mech.} \textbf{661}, 178--205.

\bibitem[]{HS91} \textsc{Hall, P. \& Smith, F. T.} 1991 On strongly nonlinear vortex/wave interactions in boundary-layer transition. \emph{J. Fluid Mech.} \textbf{227}, 641--666.
\textcolor{black}{
\bibitem[]{HS91} \textsc{Heinemann, T., McWilliams, J. C. \& Schekochihin, A. A.} 2011 Large-scale magnetic field generation by randomly forced shearing waves. \emph{Phys. Rev. Lett.} \textbf{107}, 255004.}

\bibitem[]{RHG97} \textsc{Rincon, F., Ogilvie, G. I. \& Proctor, M. R. E.} 2007 Self-sustaining nonlinear dynamo process in Keplerian shear flows. \emph{Phys. Rev. Lett.} \textbf{98}, 254502.

\bibitem[]{RHG97} \textsc{Rincon, F., Ogilvie, G. I., Proctor, M. R. E. \& Cossu, C.} 2008 Subcritical dynamos in shear flows. \emph{Astron. Nachr.} \textbf{329}, 750--761.

\bibitem[]{RHG97} \textsc{Riols, A., Rincon, F., Cossu, C., Lesur, G., Longaretti, P. -Y., Ogilvie, G. I., \& Herault, J.} 2013 Global bifurcations to subcritical magnetorotational dynamo action in Keplerian shear flow. \emph{J. Fluid Mech.} \textbf{731}, 1--45.

\bibitem[]{RHG97} \textsc{Roberts, P. H.} 1964 The stability of hydromagnetic Couette flow. \emph{Proc. Camb. Phil. Soc.} \textbf{60}, 635--651.

\bibitem[]{RHG97} \textsc{Rudiger, G.} 2003 Linear magnetohydrodynamic Taylor-Couette instability for liquid sodium. \emph{Phys. Rev. E} \textbf{67}, 046312.

\bibitem[]{SGH91} \textsc{Sakurai, T., Goossens, M., Hollweg, J. V.} 1991 Resonant behaviour of MHD waves on magnetic flux tubes. I. Connection formulae at the resonant surfaces. \emph{Solar Physics} \textbf{133}, 227--245.

\bibitem[]{SGH91} \textsc{Schmiegel, A.} 1999 \emph{Transition to turbulence in linearly stable shear flows.} \textbf{133}, PhD thesis. Philipps-Universita?t Marburg.

\bibitem[]{RHG97} \textsc{Skufca, J. D., Yorke, J. A. \& Eckhardt, B.} 2006 Edge of chaos in a parallel shear flow. \emph{Phys. Rev. Lett.} \textbf{96}, 174101.

\bibitem[]{RHG97} \textsc{Soler, R., Ruderman, M. S. \& Goossens, M.} 2012 Damped kink oscillations of flowing prominence threads. \emph{Astron. Astrophys.} \textbf{546}, A82.
\textcolor{black}{
\bibitem[]{RHG97} \textsc{Teed, R. J. \& Proctor, M. R. E.} 2017 Quasi-cyclic behaviour in non-linear simulations of the shear dynamo. \emph{Mon. Not. Roy. Astron. Soc.} \textbf{467}, 4858--4864.
}
\textcolor{black}{
\bibitem[]{RHG97} \textsc{Tobias, S. M. \& Cattaneo, F.} 2013 Shear-driven dynamo waves at high magnetic Reynolds number. \emph{Nature} \textbf{497}, 463--465.}

\bibitem[]{RHG97} \textsc{van Ballegooijen, A. A., Asgari-Targhi, M., Cranmer, S. R. \& DeLuca, E. E.} 2011 Heating of the solar chromosphere and corona by Alfv\'en wave turbulence. \emph{Astrophys. J.} \textbf{736}:3, 1--27.

\bibitem[]{RHG97} \textsc{van Veen, L. \& Kawahara, G.} 2011 Homoclinic tangle on the edge of shear turbulence. \emph{Phys. Rev. Lett.} \textbf{107}, 114501.

\item \textsc{Waleffe, F.} 1997 On a self-sustaining process in shear flows. \textit{Phys. Fluids} \textbf{9}, 883--900.

\item \textsc{Waleffe, F.} 2003 Homotopy of exact coherent structures in plane shear flows. \textit{Phys. Fluids} \textbf{15}, 1517--1534.
\textcolor{black}{
\item \textsc{Walker, J. \& Boldyrev, S.} 2017 Magnetorotational dynamo action in the shearing box. \textit{Mon. Not. Roy. Astron. Soc.} \textbf{470}, 2653--2658.}
\textcolor{black}{
\item \textsc{Walker, J., Lesur, G. \& Boldyrev, S.} 2016 On the nature of magnetic turbulence in rotating, shearing flows. \textit{Mon. Not. Roy. Astron. Soc.} \textbf{457}, L39--L43.}

\item \textsc{Wang, J., Gibson, J. F. \& Waleffe, F.} 2007 Lower branch coherent states: transition and control. \emph{Phys. Rev. Lett.} \textbf{98}, 204501.
\textcolor{black}{
\item \textsc{Willis, A. P. \& Barenghi, C. F.} 2002a A Taylor-Couette dynamo. \emph{Astron. Astrophys.} \textbf{393}, 339--343.}

\item \textsc{Willis, A. P. \& Barenghi, C. F.} 2002b Hydromagnetic Taylor-Couette flow: numerical formulation and comparison with experiment. \emph{J. Fluid Mech.} \textbf{463}, 361--375.
\textcolor{black}{
\item \textsc{Yousef, T. A., Heinemann, T., Schekochihin, A. A., Kleeorin, N., Rogachevskii, I, Iskakov, A. B., Cowley, S. C. \& McWilliams, J. C.} 2008 Generation of magnetic field by combined action of turbulence and shear. \emph{Phys. Rev. Lett.} \textbf{100}, 184501.}

\end{thebibliography}
\end{document}